\documentclass[fleqn,usenatbib]{mnras}

\usepackage{newtxtext,newtxmath}

\usepackage[T1]{fontenc}
\usepackage{siunitx}

\DeclareRobustCommand{\VAN}[3]{#2}
\let\VANthebibliography\thebibliography
\def\thebibliography{\DeclareRobustCommand{\VAN}[3]{##3}\VANthebibliography}

\usepackage{graphicx}	
\usepackage{amsmath}	

\title[Drag force and gravitational instability -- I]{The role of the drag force in the gravitational stability of dusty planet forming disc -- I. Analytical theory}

\author[Longarini et al.]{
Cristiano Longarini,$^{1}$\thanks{E-mail: cristiano.longarini@unimi.it}
Giuseppe Lodato$^{1}$,
Giuseppe Bertin$^{1}$ and 
Philip J. Armitage$^{2,3}$ \\
$^{1}$Dipartimento di Fisica, Università degli Studi di Milano, via Celoria 16, 20133 Milano, Italy\\
$^{2}$Department of Physics and Astronomy, Stony Brook University, Stony Brook, NY 11794, USA \\
$^{3}$Center for Computational Astrophysics, Flatiron Institute, New York, NY 10010, USA 
}

\date{Accepted XXX. Received YYY; in original form ZZZ}

\pubyear{2021}

\begin{document}
\label{firstpage}
\pagerange{\pageref{firstpage}--\pageref{lastpage}}
\maketitle

\begin{abstract}
Recent observations show that planet formation is already underway in young systems, when the protostar is still embedded into the molecular cloud and the accretion disc is massive. In such environments, the role of self gravity (SG) and gravitational instability (GI) is crucial in determining the dynamical evolution of the disc. In this work, we study the dynamical role of drag force in self-gravitating discs as a way to form planetesimals in early protoplanetary stages. We obtain the dispersion relation for density-wave perturbations on a fluid composed of two phases (gas and dust) interacting through the common gravitation field and the mutual drag force, and we find that the stability  {threshold} is determined by three parameters: the local dust-to-gas density ratio, the dust relative temperature and the relevant Stokes number.  {In a region of parameters space, where young protoplanetary discs are likely to be found, the instability can be \textit{dust driven}, occurring at small wavelengths. In this regime, the Jeans mass is much smaller than the one predicted by the standard gravitational instability model.} This mechanism can be a viable way to form planetary cores in protostellar discs, since their predicted mass is about $\sim10\text{M}_\oplus$.
\end{abstract}




\section{Introduction}

In young massive protoplanetary systems, the role of self-gravity is dynamically important since the disc contribution to the gravitational potential is non-negligible \citep{bertinlod} and the system may develop gravitational instability \citep{cossins,lodkr}. An example is the system Elias 2-27 \citep{perezelias,paneque}: the spectacular images of a prominent spiral structure strongly indicate that instability processes are significant, and recently its mass has been computed in a dynamical way, measuring the deviation from Keplerian rotation due to self gravity \citep{bennielias}. Such systems are characterized by the presence of a spiral pattern that can be either in a thermally saturated state or in an unstable condition \citep{lodkr}. In the first case, a stationary spiral transports angular momentum through the disc, while in the second case it fragments, potentially leading to the formation of low mass stellar companions or giant planets. In addition, spiral structure also influences gas kinematics, resulting in deviations from Keplerian rotation visible in observed velocity field \citep{hallwigg,longwig,terrywig}. While the gas behaviour in self-gravitating environments has been deeply explored, the dust behaviour has not yet been completely understood. 

Protoplanetary discs are thought to be the environments where planets form, and they are made up of gas and dust. The two components are aerodynamically coupled through a drag force, caused by the difference in velocity between them. The strength of the coupling is measured through a dimensionless parameter, the Stokes number, that is the ratio between the stopping time of dust particles and the dynamical time. Solid particles with $\text{St}<<1$ are strongly coupled to the gas, conversely the ones with $\text{St}>>1$ are weakly coupled. In  {the Epstein regime}, it is true that $\text{St}\propto a/\Sigma_g$, where $a$ is the dust grain size and $\Sigma_g$ is the gas surface density. The main effect of the aerodynamical coupling between gas and dust is the so-called ``radial drift'': the gas has a sub-Keplerian azimuthal velocity, due to its pressure gradient, and a radial velocity given by viscous effects. In contrast, the dust velocity is almost Keplerian, being a pressureless and inviscid fluid, at a first approximation. Thus, solid particles experience an azimuthal headwind, and the resulting drag force
tends to slow the dust down and lets it migrate toward the central object. The maximum radial drift occurs for dust particles with $\text{St}=1$: this phenomenon has dramatic effects in the context of planet formation, since its timescale is very short, of the order of $\sim\SI{100}{yr}$, preventing dust growth.

Investigating dust dynamics in protostellar discs is essential to understand planet formation, since dust is the fundamental constituent of planetary cores. The large number of substructures, such as rings and gaps, in class II objects $(\sim 1-5 \text{Myr})$ is often explained with the presence of protoplanets embedded in the disc \citep{dharp1,dsharp2}. The mass of these objects can be inferred in different ways, such as morphological imprints \citep{morph1, morph2}, kinematic signatures \citep{kinkpinte,kinkbollati} or hydrodynamical modelling \citep{claHD,benniDS}. It turns out that many of class II systems are hosting planets with several Jupiter masses, as in the case of PDS 70 \citep{kepplerPDS,kepplerPDS19,claPDS,facchiniPDS}, in which there are two giant planets with masses larger than 5 Jupiter masses. This necessarily implies that planet formation is already underway in young systems (class I objects), when the disc still interacts with the envelope. An example is the system IRS 63 \citep{cox}, a young $(\sim 0.5\text{Myr})$ protoplanetary disc that shows four annular structures carved by a system of protoplanets. 

The classical theory of planet formation is the Core Accretion model \citep{CA1,CA2} (hereafter CA), and it is focused on dust dynamics. Four stages can be identified, according to the grains' size and Stokes number: the first one is called ``dust growth" \citep{dustgrowth} as dust coagulates from micron-size to centimetre-size through microphysical mechanisms. In this stage, the Stokes number is below unity, and it increases according to the dust grains' size. The second stage is called ``planetesimal formation" and it has not yet been completely understood. As a matter of fact, it is not clear how dust grows from centimetre to kilometre size, forming the so-called planetesimals. Indeed, during this stage the Stokes number is $\sim 1$ and thus radial drift effect is maximum. For this reason, the presence of a mechanism that stops radial drift is fundamental for planet formation, otherwise dust growth would stop \citep{metresized}. 

The third step is called ``collisional growth": through scattering and collisions, planetesimals reach planetary cores dimensions. Collisions are fostered by a mechanism called ``gravitational focusing", thanks to which the geometrical cross-section increases because of gravitational attraction. This effect promotes the growth of bigger planetesimals and for this reason it is called ``oligarchic growth" \citep{kokubo}. The last stage is ``core accretion", in which planetary cores accrete gas forming their atmospheres and, under certain conditions, become gaseous giants \citep{coreinst}. Although this model currently provides the most popular explanation for the formation of planets, it has been shown that formation timescales may be anywhere up to $\sim \SI{10}{Myr}$, exceeding typical disc lifetimes \citep{NOCA1}, specifically in the case of giant planets in the outer disc. For example, the formation of systems as HD100546 \citep{davideHD} or CIDA 1 \citep{CIDA1, pietro} is particularly challenging for CA theory.

An alternative to the CA model is the Gravitational Instability scenario \citep{bossGI}: unstable regions of the disc may collapse into  {gravitationally bound} fragments, forming giant gaseous planets or, more likely, low mass stellar companions \citep{kratter1}. The mass of the fragment created by GI is the Jeans mass, and for typical protoplanetary discs its value is between $1-10\text{M}_\text{J}$. However, this is the initial mass of the object: indeed, the fragment would start accreting material belonging to the accretion disc, increasing its mass and likely becoming a brown dwarf \citep{kratter2}.

As we have just pointed out, current models of planet formation are not able to explain the formation of \emph{planetesimals}, CA because of radial drift and GI because of too large value of Jeans mass. A potential path to solve this issue is the streaming instability \citep{stream1,stream2,stream3}, and it can be qualitatively described as follows. Since the dust-to-gas ratio in the mid-plane is higher, the back-reaction from dust particles onto the gas is stronger, and therefore the azimuthal difference of speed is reduced, making the headwind on the particles weaker. If a dust overdensity was present, it would perturb the system, causing a stronger backreaction, and so a reduced radial drift. It is clear that this is a runaway situation, as the initial dust overdensity increases more and more due to the new material drifting inward from outer orbits and stopping into the clump: the consequence is an exponential growth of clumps, since  growth rates are slower than dynamical but faster than radial drift timescales. The effect of streaming instability is maximum for marginally coupled dust particles ($\text{St}\sim1$) and for high dust-to-gas ratio. Observationally speaking, overdensity formation via streaming instability appears to be consistent with recent multi-wavelength ALMA observations in Lupus star forming region \citep{scardoni,tazz1,tazz2}.

Nevertheless, a step forward can be made in GI planet formation scenario. Indeed, it is possible to consider the role of a second component, i.e. dust in protostellar case: gravitational instability of a two fluid system has been investigated in the context of galactic dynamics \citep{jogsolomon,bertinromeo}, and it has been shown that even a small amount of a second cold component can make the system more unstable. As for protoplanetary discs, the coupled effect of drag and GI has been numerically studied by \citet{rice04,rice06}, who find that solid cores can form rapidly in the outer disc from dust concentration in the spiral structure of a non-fragmenting gravitationally unstable disc. Indeed, gas spiral arms are maxima of pressure and minima of gravitational potential, hence dust particles are efficiently  {concentrated inside them, enhancing their density and collision rate and eventually accelerating planetesimals formation through direct gravitational collapse.} In this picture, an important parameter is the dust velocity dispersion $c_d$: in gravitationally unstable disks, \citet{boothclarke} found that $c_d$ increases with Stokes number, and it is minimum for $\text{St}\sim1$, since the combined effect of drag and gravitational force is maximum.  {This trend was also confirmed by \cite{baehr} through 3D shearing box simulations.} 

In this paper, we provide an analytical framework to study the gravitational stability of a dust and gas disc, including the effect of drag force.  {By taking into account the dust, there is a region of parameters space where the gravitational instability is driven by this component. In this context, the role of drag force is crucial, making the system transition from the one fluid to the two fluid limit. The parameter that controls this transition is the Stokes number.} 

First of all, we recall one-component and two-component fluid models for gravitational instability and we present the dispersion relation including the drag force, studying the marginal stability curve and discussing the parameters' space \ref{S3}. In section \ref{S4} we apply drag gravitational instability to protostellar discs, and we discuss the implications in terms of planet formation, presenting some limits  {and comparing our results with previous works.} Finally, in section \ref{S5} we draw the conclusions of the work.

\section{One and two fluids instability}\label{S3}

\begin{figure}
	\includegraphics[width=\columnwidth]{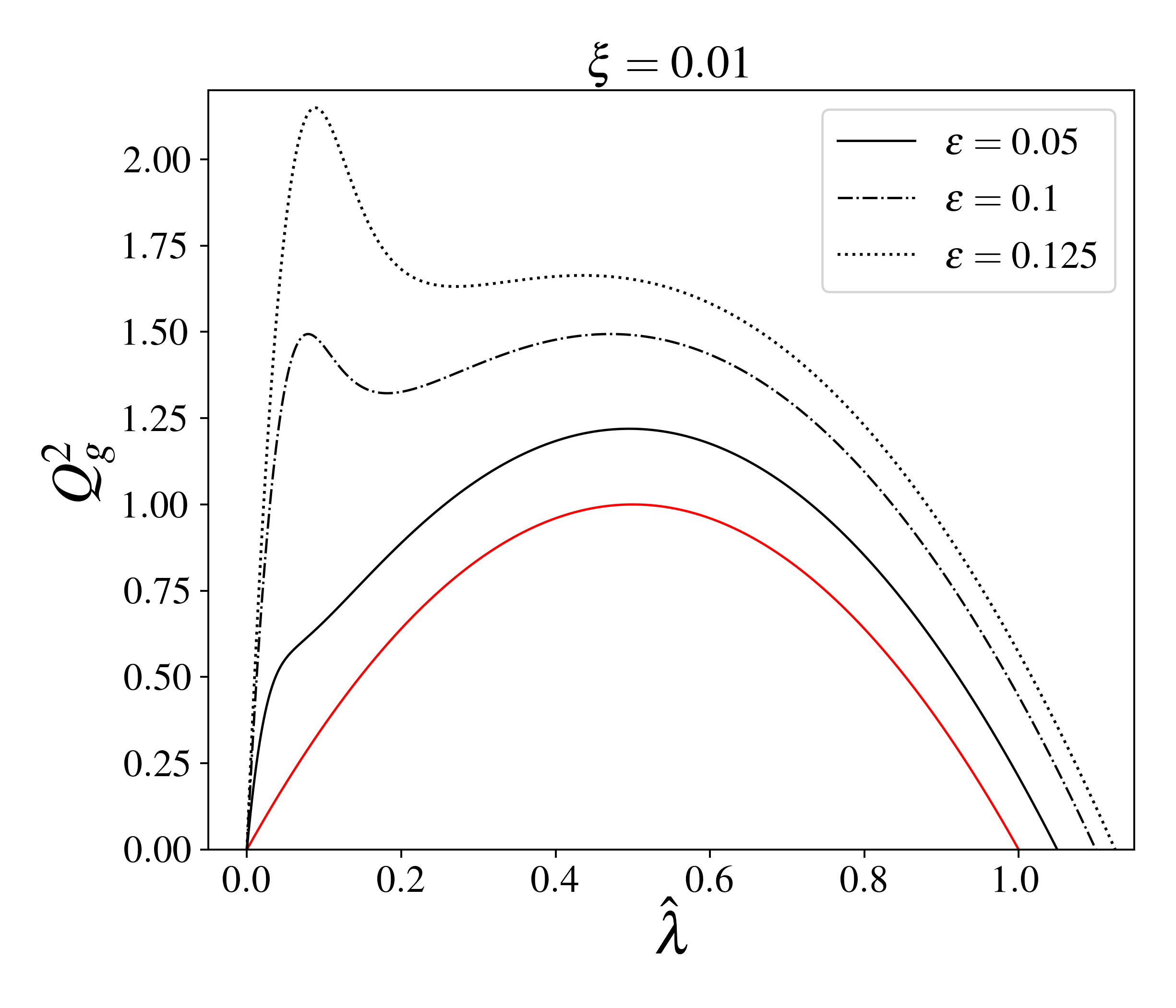}
    \caption{Marginal stability curve of the one-component fluid model (red line) and of the two-component fluid one (black lines) for $\xi = 0.01$ and different values of $\epsilon$. It can be seen that for $\epsilon<\sqrt{\xi}$ the ``hot'' peak is dominant, while for $\epsilon>\sqrt{\xi}$ the ``cold'' peak at smaller wavelengths is higher.}
    \label{marginal1_2}
\end{figure}

In this work, we study the gravitational instability of an axisymmetric protoplanetary disc. We first consider a thin disc composed of a single fluid, with surface density $\Sigma$ and sound speed $c$,  differentially rotating with angular frequency $\Omega$ and epicyclic frequency $\kappa$. Under the hypothesis of an axisymmetric disc, all of these quantities are a function of radius $r$. The dispersion relation for linear, tightly-wound density waves is
\begin{equation}
    (\omega-m\Omega)^2 = c^2k^2 - 2\pi G \Sigma|k| + \kappa^2,
\end{equation}
where $\omega$ is the perturbation frequency and $m/r$ is the azimuthal wavenumber of the perturbation, and they are constant over the radial extent of the disc \citep{safr60}. The theory was first developed in the context of galactic dynamics, and it is the foundation of the density wave theory for spiral galaxies \citep{Linshu,toomre64}.  More recently, an extensive characterization of GI in protoplanetary system has been presented by \cite{cossins}. For axisymmetric perturbations ($m=0$), the instability threshold is determined by the parameter
\begin{equation}\label{toomreq}
    Q = \frac{c\kappa}{\pi G \Sigma};
\end{equation}
when $Q>1$, the system is stable at all wavelengths, if $Q<1$ there is an interval of unstable wavelengths and $Q=1$ is the marginally unstable case. The most unstable wave is characterized by having the following wavenumber
\begin{equation}
    k_J = \frac{\pi G \Sigma}{c^2},
\end{equation}
or equivalently wavelength
\begin{equation}
    \lambda_J = \frac{2 \pi }{k_J} = \frac{2c^2}{G \Sigma}.
\end{equation}
$k_J$ and $\lambda_J$ are respectively the most unstable wavenumber and wavelength, often called Jeans wavenumber and wavelength. Finally, it is possible to define the Jeans mass, that is the mass enclosed in a Jeans wavelength \citep{lodkr}
\begin{equation}
    M_J = \Sigma \lambda_J^2 = \frac{4c^4}{G^2\Sigma}.
\end{equation}
For protoplanetary discs, the thin disc approximation requires the approximate relation
\begin{equation}
    \frac{c}{v_k} = \frac{H}{r},
\end{equation}
where $v_K = \sqrt{GM_\star /r}$, $H$ can be obtained by the hydrostatic equilibrium, and it is given by $H= c/\Omega$ and $H/r$ is the aspect ratio, that is usually taken as $H/r\simeq0.1$. By using this identity, the Jeans mass can be written as
\begin{equation}
    M_J = 4\pi \left(\frac{H}{r}\right)^4\frac{M_\star}{M_d} M_\star = 10^{-2}\left(\frac{H/r}{0.1}\right)^4\left(\frac{M_\star/M_d}{10}\right)M_\star;
\end{equation}
this means that for a star mass of the order of the solar mass, the typical Jeans mass in protoplanetary discs is of the order of $10M_\text{jup}$, where $M_\text{jup}$ is the Jupiter mass, too high to form a protoplanet.

The instability threshold can be studied by inspection of the marginal stability curve, which is a plot of the squared stability parameter $Q^2$ as a function of the dimensionless perturbation wavelength, by imposing the marginal stability condition $\omega-m\Omega=0$. Above this curve, the system is stable, while below it is unstable, and the maximum of the curve is the critical value of $Q$. To draw the marginal stability curve, we define dimensionless Doppler-shifted perturbation frequency and wavelength
\begin{equation}
    \nu = \frac{\omega-m\Omega}{\kappa},\quad \hat{\lambda} = \frac{\kappa^2}{2\pi G \Sigma k};
\end{equation}
with these identities, the marginal stability curve is given by $\nu = 0$, and it reads 
\begin{equation}
    Q^2 = 4\left(\hat{\lambda}-\hat{\lambda}^2\right).
\end{equation}

It is possible to move toward a more realistic description of protoplanetary discs, which are
composed of two main elements, gas and dust. The gravitational stability for a multi-phase fluid is different. The two components can be described as two different fluids with surface density $\Sigma_g$, $\Sigma_d$ and sound speed $c_g$ and $c_d$, where the subscript $g$ refers to the gas and $d$ to the dust. In the context of protoplanetary discs, gas component is more abundant $\Sigma_g > \Sigma_d$ and hotter $c_g > c_d$. It is possible to define two parameters that measure these properties: the first one is the local relative abundance of dust 
\begin{equation}
    \epsilon = \frac{\Sigma_d}{\Sigma_g},
\end{equation}
often called dust-to-gas ratio; the second one is the local dust relative temperature
\begin{equation}
    \xi = \left(\frac{c_d}{c_g}\right)^2.
\end{equation}
As already pointed out, for a typical protoplanetary disc both $\epsilon$ and $\xi$ are smaller than unity. Clearly, the one-component fluid model limit is given by the condition $\epsilon \to 0$.

\cite{kato72} first proposed that the presence of a second cold component can trigger gravitational instability at small scales: \cite{jogsolomon} and \cite{bertinromeo} proposed a quantitative analysis of this behaviour, by computing the dispersion relation and the marginal stability curve. In this model, the two components are coupled to each other only through the common gravitational field. The dispersion relation for the two-component fluid model reads
\begin{equation}\label{JS}
    \omega^4 -\omega^2(\alpha_g+\alpha_d) + (\alpha_g\alpha_d -\beta_g\beta_d) = 0,
\end{equation}
with $\alpha_i = \kappa^2 + c_i^2k^2 - 2\pi G \Sigma_{i}|k|$ and $\beta_i = 2\pi G \Sigma_{i}|k|$.
and the marginal stability curve is given by
\begin{equation}\label{MSC_JS}
\begin{array}{l}
  Q_{g}^{2}=\frac{2 \hat{\lambda}}{\xi}\left[(\epsilon+\xi)-\hat{\lambda}(1+\xi)+\right. \\
  \left.\sqrt{\hat{\lambda}^{2}(1-\xi)^{2}-2 \hat{\lambda}(1-\xi)(\epsilon-\xi)+(\epsilon+\xi)^{2}}\right],
 \end{array}
\end{equation}
where the parameter $Q_g$ is defined as in Eq. \ref{toomreq} but in terms of only the properties $\Sigma_g$ and $c_g$ of the gas component. A comparison between the marginal stability curve of the one-component fluid model and the two-component one is shown in figure \ref{marginal1_2}. The profile of $Q_g$ may exhibit two peaks, one arising from instability in the gas component, at intermediate wavelengths, and one at smaller wavelengths, dominated by the cold component. This second peak emerges when the dust is sufficiently abundant and cold: \cite{bertinromeo} found that there is a transition from gas to dust driven instability when $\epsilon > \sqrt{\xi}$. In the two-component fluid model, the Jeans length is defined as the wavelength at which $Q_g^2$ has its maximum, so when instability is dust driven, the Jeans length is smaller compared to the gas driven case. As in the previous case, the theory was first developed in the context of spiral galaxies, where the two fluids are stars and gas. In the galactic context, stars are the hotter and more abundant component, in turn the gas is the secondary colder component.

In protoplanetary discs, gas and dust are aerodynamically coupled through a drag force: its role is essential in determining the dynamical evolution of the system, and so far it has been neglected. 

In the following subsection, we present our results concerning the role of the drag force between gas and dust in the gravitational stability of density waves in a protplanetary disc. We obtain the dispersion relation for a two-component fluid model, including aerodynamical coupling: firstly, we neglect the dust backreaction, then we consider it and we evaluate its effect. The details of the calculations can be found in appendix \ref{appa}. Then, we obtain the marginal stability curve, and we compare it with that of the one-component and two-component fluid models considered earlier in this section.

\subsection{Gravitational and aerodynamical coupling between gas and dust}
In this analysis, we consider a gas and dust disc, where the two components are coupled through both gravitational and drag force. The gravitational interaction is described by the Poisson equation, that for this system is
\begin{equation}\label{poisson}
    \nabla^2\Phi = 4\pi G \delta(z) (\Sigma_g+\Sigma_d),
\end{equation}
where $\Phi$ is the total gravitational potential,  $\Sigma_g,\Sigma_d$ are the surface densities of gas and dust respectively and $\delta(z)$ is the Dirac Delta function. Equation (\ref{poisson}) is telling us that both gas and dust contribute to the total gravitational potential. As for the aerodynamical coupling, it appears in the Euler equations
\begin{equation}\label{drag1eq}
    \partial_t \mathbf{v}_g + (\mathbf{v}_g \cdot \nabla)\mathbf{v}_g = - \nabla (\Phi + h_g) + \frac{1}{\Sigma_g}\mathbf{F}_d ,
\end{equation}
\begin{equation}
    \partial_t \mathbf{v}_d + (\mathbf{v}_d \cdot \nabla)\mathbf{v}_d = -\nabla (\Phi + h_d) - \frac{1}{\Sigma_d}\mathbf{F}_d ,
\end{equation}
where $\mathbf{v}_g,\mathbf{v}_d$ are gas and dust velocity vectors, $h_g,h_d$ are gas and dust enthalpies\footnote{The enthalpy is related to the sound speed: $\text{d}h = c^2 \text{d}\Sigma/\Sigma$.} and $\mathbf{F}_d$ is the drag force per unit surface, defined as
\begin{equation}
    \mathbf{F}_d = \frac{\Sigma_d}{t_s} (\mathbf{v}_d-\mathbf{v}_g),
\end{equation}
where $t_s$ is the stopping time of dust particles, i.e. the time in which drag modifies the relative velocity significantly. By these definitions, it is evident that the effect of the drag onto the gas component is smaller than the dust one of a factor $\epsilon = \Sigma_d / \Sigma_g$, the so-called ``dust to gas ratio'', that is considered to be $\epsilon \sim 0.01$ from ISM abundances \citep{dusttogas}. Because of the small value of $\epsilon$,  we firstly neglect the effect of the backreaction (i.e. we neglect the last term in Eq. \ref{drag1eq}): this allows us to deal with simpler algebra; then, we add it, and we evaluate its effect.

\subsection{Instability without backreaction}\label{nbr}

Here, we record the dispersion relation $D_\text{nbr}(\omega, k)$ without taking into account the backreaction. To do so, we perform a first order perturbation analysis of the fluid equations ( for an outline of the derivation, see Appendix \ref{appa}). We find a fifth order equation with complex coefficients, that reads
\begin{equation}\label{disp_nbr}
\begin{array}{ll}
    D_\text{nbr}(\omega,k)=-i\omega^5 + \frac{2\omega^4}{t_s} + i \omega^3\left(\alpha_g+\alpha_d+\frac{1}{t_s^2}\right) +\\- \frac{\omega^2}{t_s}\left(2\alpha_g+\alpha_d-\beta_d-\kappa^2\right) +\\ -i\omega\left[\alpha_g\alpha_d -\beta_g\beta_d +\frac{1}{t_s^2}(\alpha_g-\beta_d)\right] +\\ +\frac{1}{t_s}\left[\alpha_g\alpha_d-\beta_g\beta_d-\kappa^2(\alpha_g-\beta_d)\right]=0,
\end{array}
\end{equation}
where $\alpha_i = \kappa^2 + c_i^2k^2 -2\pi G \Sigma_i |k|$ and $\beta_i = 2\pi G \Sigma_i|k|$, in which the subscript $i$ takes on the values g or d to denote which species the various quantities refer to. The dispersion relation can be divided into two parts, one that contains the drag coupling and one ``drag-free'', that corresponds to the two-component fluid model one 
\begin{equation}
    D_\text{nbr}(\omega, k) = -i\omega D_\text{2f}(\omega,k) + \frac{1}{t_s}D_\text{drag}(\omega,k),
\end{equation}
with $D_\text{2f}$ given by the left-hand-side of Eq. (\ref{JS}) and
\begin{equation}
\begin{array}{ll}\label{dispdrag}
    D_\text{drag}(\omega,k) = 2\omega^4 + \frac{i\omega^3}{t_s} - \omega^2(2\alpha_g+\alpha_d-\beta_d-\kappa^2) +\\
    - \frac{i\omega}{t_s} (\alpha_g - \beta_d) + \left[\alpha_g\alpha_d-\beta_g\beta_d -\kappa^2(\alpha_g-\beta_d)   \right].
\end{array}
\end{equation}
One should note that in the limit of weak aerodynamical coupling, $t_s>>\kappa^{-1}$, where $\kappa^{-1}\simeq\Omega^{-1}$ is the dynamical time of the system, the dispersion relation reduces to the two-component fluid model one.

In order to compare this result with one and two-component fluid model, we compute the marginal stability curve. In this case, there are three parameters that determine the instability: $\epsilon$ and $\xi$, defined as before, and the Stokes number $\text{St} = t_s\kappa$, that measures the strength of the aerodynamical coupling. Since the dispersion relation has complex coefficients, the instability threshold is given by $\text{Im}(\omega) = 0$. 

We first study the high drag limit $\text{St}=t_s\kappa<<1$: in this case, Eq. \ref{dispdrag} is dominated by the two imaginary terms and we neglect the others, so we can get analytically the marginal stability curve. As for the general case, we obtain numerically the marginal stability curve by imposing the imaginary part of the roots of Eq. (\ref{disp_nbr}) to be zero, and we find that it is well reproduced by the high drag approximation; this is also true when we take into account the backreaction.

\begin{figure*}
	\includegraphics[scale=0.5]{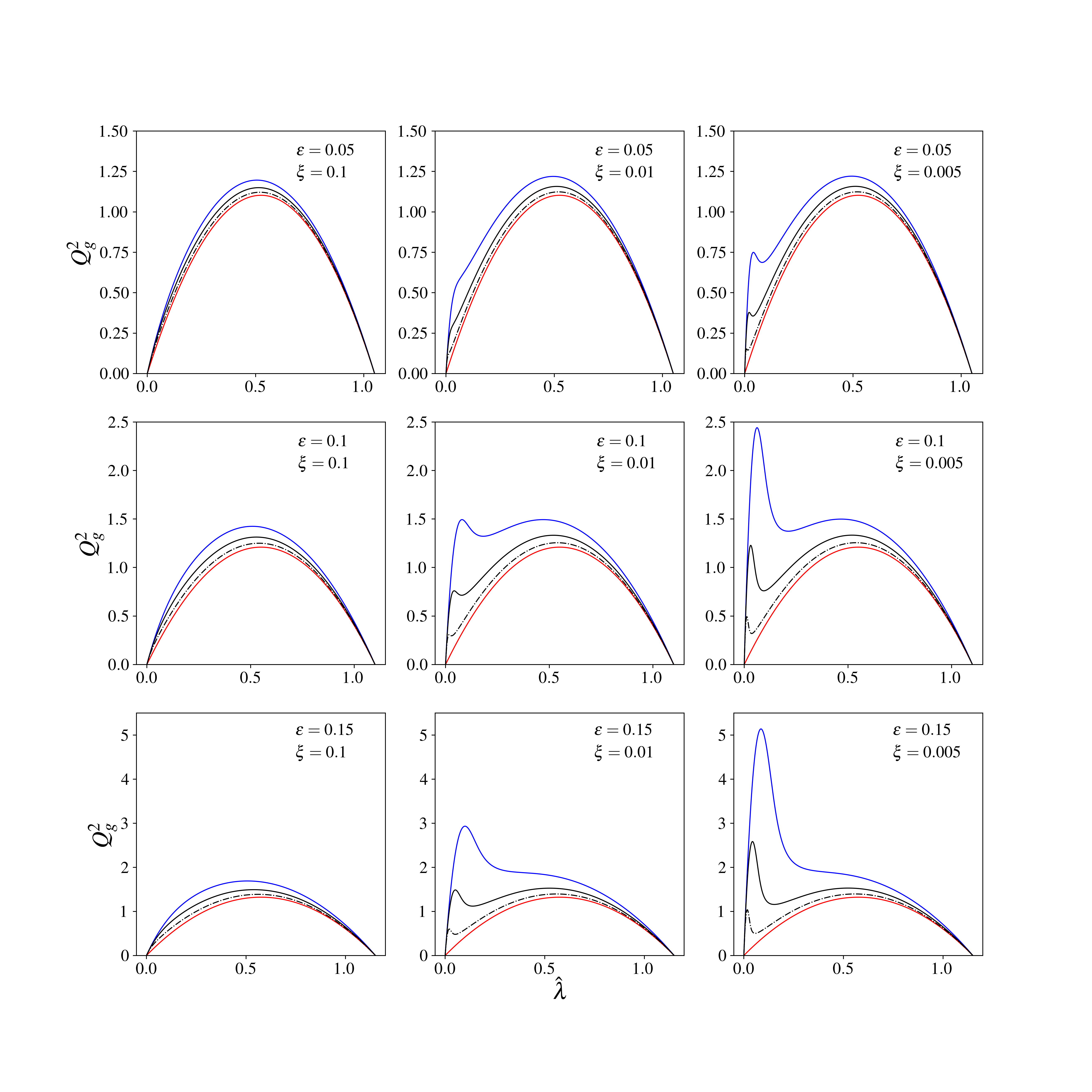}
    \caption{Examples of marginal stability curves for different values of $\epsilon$, $\xi$ and St. The blue line is the two-component fluid model without drag force, the red line is one-component fluid model and the black lines represents the two-component fluid model with drag force, without taking into account the backreaction; the solid line corresponds to $\text{St} = 1$ and the dashed line $\text{St}=0.5$.}
    \label{general_case}
\end{figure*}

In the high drag regime, the dispersion relation has the following form
\begin{equation}\label{hidrag}
    \omega^4 - \omega^2 \left( {\alpha}_g+ {\alpha}_d + \frac{1}{t_s^2}\right) + \left[ {\alpha}_g {\alpha}_d -  {\beta}_g {\beta}_d + \frac{1}{t_s^2}( {\alpha}_g- {\beta}_d)\right]=0.
\end{equation}
To write the marginal stability curve, we set $\omega = 0$ in Eq. \ref{hidrag}, we write it in a dimensionless form, and we solve for $Q_g^2$, obtaining
\begin{equation}\label{MSC_NBR}
\begin{array}{l}
     Q_g^2 = \frac{2\hat{\lambda}}{\xi}\left\{(\epsilon+\xi) -\hat{\lambda}(1+\xi+\text{St}^{-2}) +\right.\\ \left. + \sqrt{\left[\hat{\lambda}(1+\xi+\text{St}^{-2}) - (\epsilon+\xi)\right]^2+ }\right.\\
     \left. \overline{-4\xi\left[\hat{\lambda}^2\left(1+\text{St}^{-2}\right)-\hat{\lambda}\left(1+\epsilon + (1+\epsilon)\text{St}^{-2}\right)\right]} \right\}.
\end{array}
\end{equation}
It is important to note that for $\text{St}\to\infty$, Eq. (\ref{MSC_NBR}) reduces to (\ref{MSC_JS}) since drag force is negligible. On the contrary, when $\text{St}\to0$, the system behaves as a single-component fluid with $\Sigma = \Sigma_g+\Sigma_d$ Figure \ref{general_case} shows marginal stability curves for different values of $\epsilon$, $\xi$ and St compared to the one and two-component fluid models.
The role of the drag force is to connect with continuity the one and two fluid instability. We define $Q^2_\text{1f}$, $Q^2_\text{2f}$ and $Q^2_\text{D}$ as the marginal stability curve of the one-component fluid model\footnote{In the one-component fluid model, the surface density is given by $\Sigma = \Sigma_g + \Sigma_d$}, of the two-component fluid model and of the drag model; once we fix $\epsilon$ and $\xi$, the following condition is always respected
\begin{equation}
    Q^2_\text{1f} \leq Q^2_\text{D}(\text{St}) \leq Q^2_\text{2f}.
\end{equation}

\begin{figure}
	\includegraphics[scale=0.475]{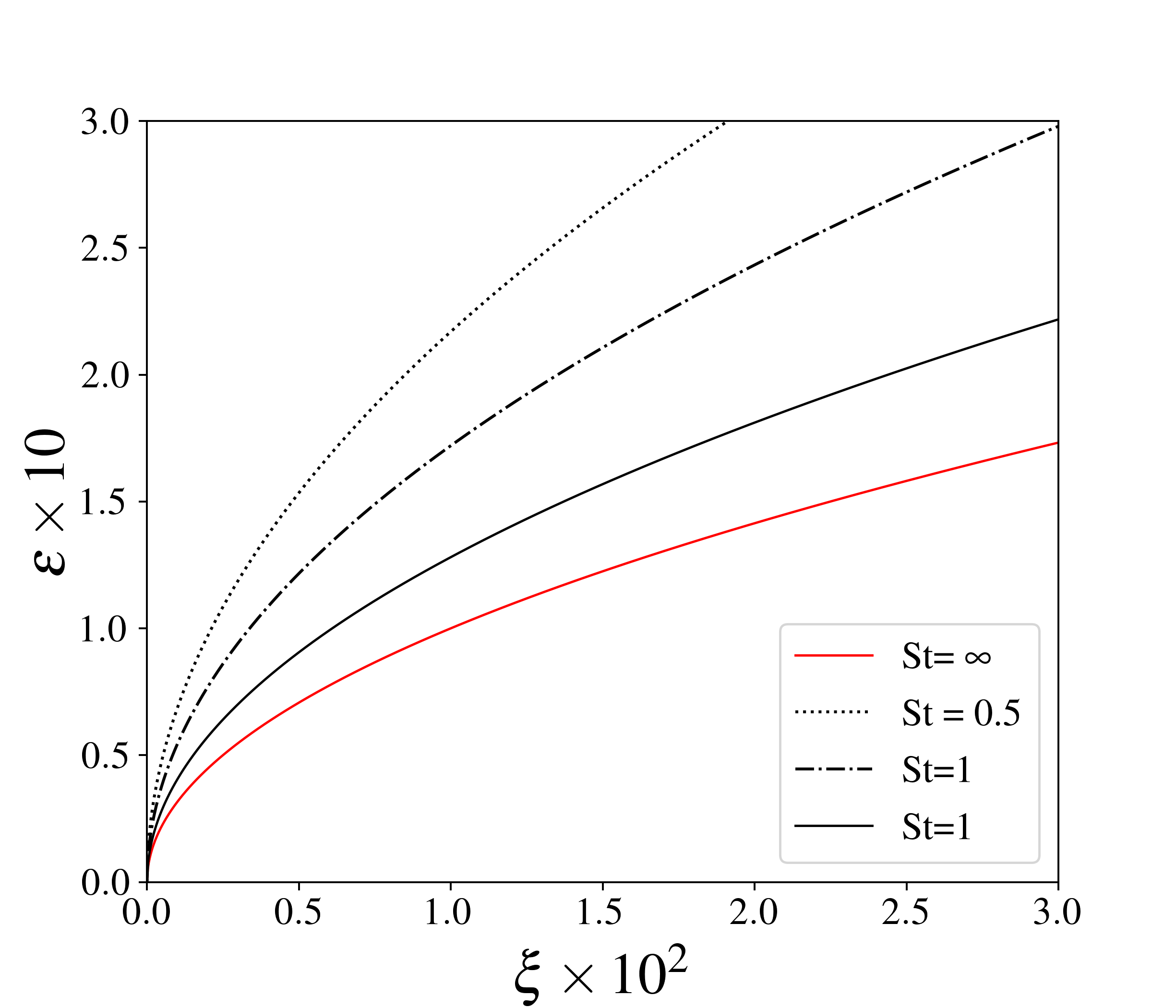}
    \caption{Transition curves for different Stokes number in the $(\xi,\epsilon)$ diagram obtained by Eq. (\ref{transition}). For a chosen Stokes number, in the region above the transition curve the instability is dust-driven, while below is gas-driven. The $\text{St}\to\infty$ case recovers the well known result $\epsilon = \sqrt{\xi}$.}
    \label{limit}
\end{figure}


As we have already pointed out, \citet{bertinromeo} found that the transition from gas to dust driven instability happens when $\epsilon > \sqrt{\xi}$: when we take into account the drag force, this relation changes. Indeed, when $\text{St}\to0$, gas and dust are strongly coupled, and they can be considered as one fluid:  {in this case, physically speaking, we expect that the velocity dispersion is the same ($\xi=1$), and thus the instability is gas driven.  {This is just an approximation: in general, the thermal velocity of small dust particles is set by Brownian motions. Modelling this phenomenon is not the purpose of this work, hence we make the approximation that for $\text{St}\to 0$ , $c_d=c_g$.} Under a mathematical point of view, even though $\text{St}\to0$, if $\epsilon\neq0$, there is always a value of $\xi$ under which the instability becomes dust driven.} Conversely, when $\text{St}\to\infty$, the system tends to the two-component fluid model, where the drag interaction is not taken into account, and thus the transition condition is the same of \citet{bertinromeo}. Now, we want to generalize the transition condition taking into account the role of the drag force: we argue that the value of $\epsilon$ at which the transition from gas to dust driven instability occurs can be written as
\begin{equation}
    \epsilon_\text{tr} = f(\text{St})\sqrt{\xi},
\end{equation}
where $f$ is a function for the Stokes number. This function must respect two conditions
\begin{equation}
    \lim_{\text{St} \to \infty} f(\text{St}) = 1, \quad \lim_{\text{St}\to 0} f(\text{St}) = \infty,
\end{equation}
in order to recover one and two fluid limits. For simplicity, we hypothesize that $f$ has the following form
\begin{equation}\label{transition}
    f(\text{St}) = 1 + a\text{St}^{b},
\end{equation}
and we found that the two best fit coefficients are $a = 0.72$, $b =- 1.36 $.  Figure \ref{limit} shows the transition curves for different values of the Stokes number in the $(\xi,\epsilon)$ diagram: for a chosen St, in the region above the curve the instability is dust-driven, while below is gas-driven.

\subsection{Instability with backreaction}
Now we follow the same path as before, but taking into account the backreaction. We start from the same hypotheses of section \ref{nbr}, and we get the dispersion relation $D_\text{br}(\omega,k)$ (an outline of the derivation is given in Appendix \ref{appa})
\begin{equation}
\begin{array}{ll}
    -i\omega^5 + \frac{2\omega^4}{t_s}(1+\epsilon) + i \omega^3\left[\alpha_g+\alpha_d+\frac{1}{t_s^2}(1+2\epsilon+\epsilon^2)\right] +\\- \frac{\omega^2}{t_s}\left[(2\alpha_g+\alpha_d-\kappa^2)(1+\epsilon)-\beta_d\right] +\\ -i\omega\left[\alpha_g\alpha_d -\beta_g\beta_d +\frac{1}{t_s^2}(\alpha_g-\beta_d+\right.\\\left.+ \epsilon({\alpha}_{g}+{\alpha}_{d}- {\beta}_g- {\beta}_d) + \epsilon^2({\alpha}_{d}-{\beta}_g))\right] +\\ +\frac{1}{t_s}\left[(\alpha_g\alpha_d-\beta_g\beta_d)(1+\epsilon)-\kappa^2\left(\alpha_g-\beta_d - \epsilon({\alpha}_d - {\beta}_g )\right)\right]=0.
\end{array}
\end{equation}
As before, we write the dispersion relation in the high drag regime, given by the condition $\text{St} = t_s\kappa << 1$

\begin{equation}
\begin{array}{ll}
    \omega^4 - \omega^2\left[\alpha_g+\alpha_d+\frac{1}{t_s^2}(1+2\epsilon+\epsilon^2)\right] + \left[\alpha_g\alpha_d -\beta_g\beta_d + \right.\\\left. +\frac{1}{t_s^2}(\alpha_g-\beta_d+ \epsilon({\alpha}_{g}+{\alpha}_{d}- {\beta}_g- {\beta}_d) + \epsilon^2({\alpha}_{d}-{\beta}_g))\right].
\end{array}
\end{equation}
We obtain the marginal stability curve by setting $\omega=0$ in the last equation
\begin{equation}\label{MSC_BR}
\begin{array}{l}
     Q_g^2 = \frac{2\hat{\lambda}}{\xi}\left\{(\epsilon+\xi) -\hat{\lambda}(1+\xi+\text{St}^{-2} +f_1) +\right.\\ \left. + \sqrt{\left[\hat{\lambda}(1+\xi+\text{St}^{-2}+f_1) - (\epsilon+\xi)\right]^2+ }\right.\\
     \left. \overline{-4\xi\left[\hat{\lambda}^2\left(1+\text{St}^{-2}+f_2\right)-\hat{\lambda}\left(1+\epsilon + (1+\epsilon)\text{St}^{-2}+f_3\right)\right]} \right\},
\end{array}
\end{equation}
where
\begin{equation}
    f_1(\epsilon,\text{St}) = \frac{\epsilon}{\text{St}^2}(1+\xi+\epsilon\xi),
\end{equation}
\begin{equation}
    f_2(\epsilon,\text{St}) = \frac{\epsilon}{\text{St}^2}(2+\epsilon),
\end{equation}
\begin{equation}
    f_3(\epsilon,\text{St}) = \frac{\epsilon}{\text{St}^2}[2(1+\epsilon)+\epsilon(\epsilon+1)],
\end{equation}
are three correction factors.

\begin{figure*}
	\includegraphics[scale=0.465]{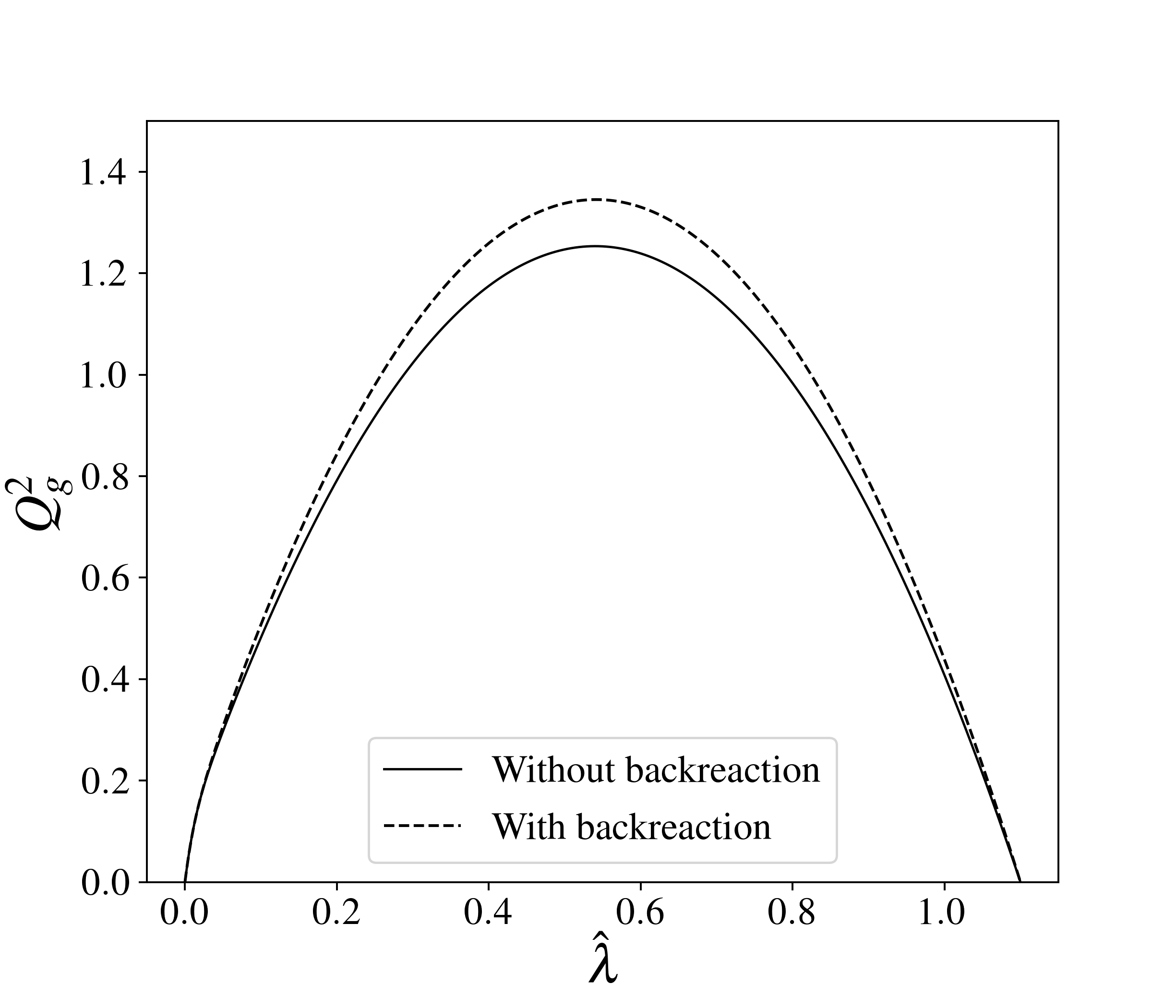}
	\includegraphics[scale=0.465]{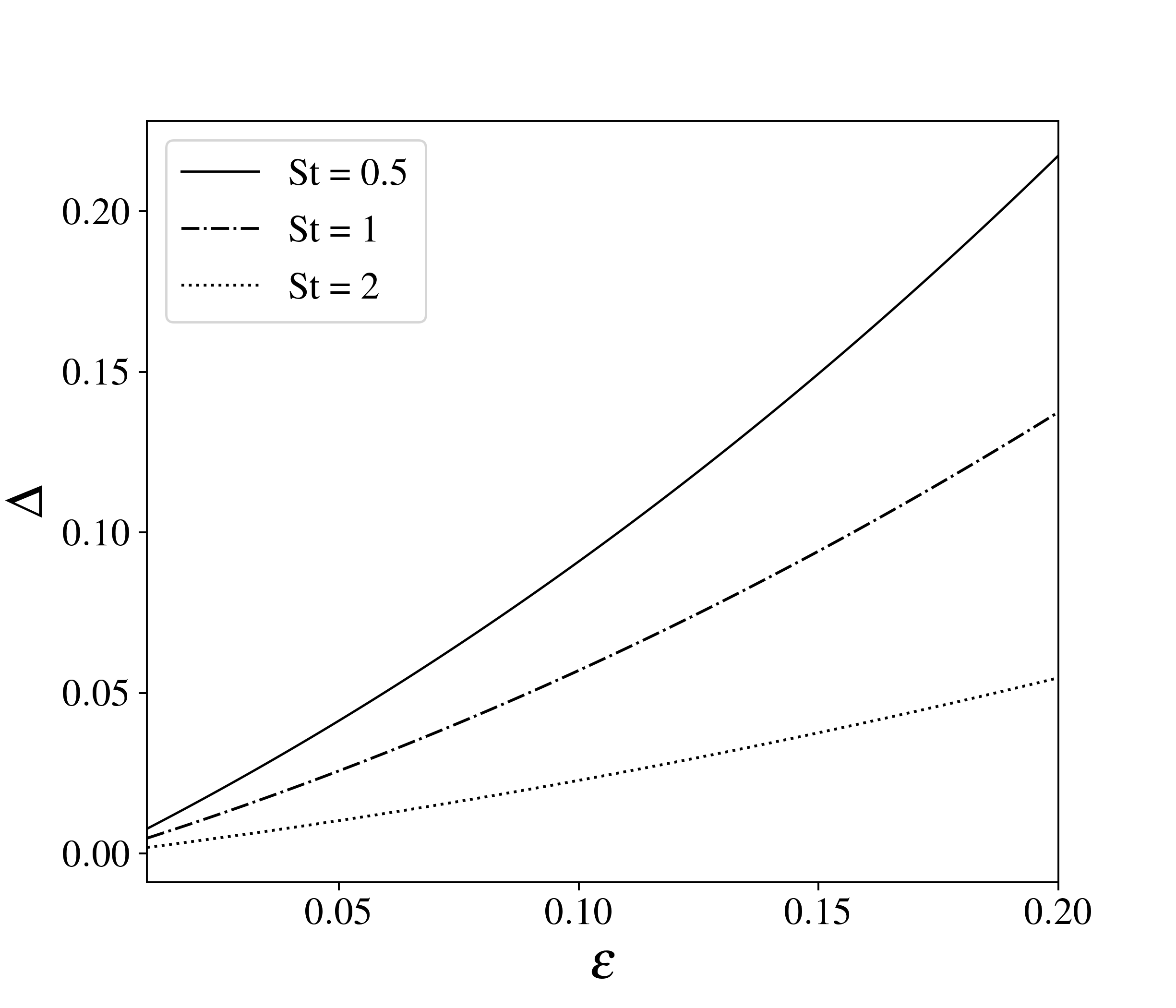}
    \caption{The effect of the backreaction onto the marginal stability curve. Left panel: marginal stability curve with and without backreaction, for $\epsilon = 0.1$, $\xi = 0.05$ and $\text{St} = 0.5$. The effect of the backreaction is maximum in correspondence to the gaseous peak. Right panel: the quantity $\Delta$ that quantifies the effect of the backreaction as a function of dust-to-gas ratio, for different values of Stokes number and for $\xi =0.05 $.}
    \label{differences}
\end{figure*}

Physically speaking, the backreaction is the effect of the drag force onto the gas component: hence we expect to see differences at intermediate wavelengths, where the instability is gas-driven. The left panel of figure \ref{differences} shows a comparison between the marginal stability curve with and without the backreaction: as expected, the height of the gaseous peak is different. In particular, when we take into account the backreaction, the system is more unstable i.e. the gaseous peak is higher. This is in agreement with what found in secular GI: backreaction makes the secular GI operational at intermediate wavelengths \citep{SGI14}. In addition, the effect of the backreaction is stronger when the dust-to-gas ratio is bigger and the Stokes number is smaller: in order to measure the effects of the backreaction, we define a new quantity
\begin{equation}
    \Delta = \text{Max}\left[\left|Q^2_\text{br} - Q^2_\text{nbr}\right|\right],
\end{equation}
shown in right panel of figure \ref{differences}. It can be clearly seen that the effect of the backreaction is small, even for extreme cases (high dust-to-gas ratio and low Stokes number): from now on, we will use the dispersion relation without backreaction, for computational convenience.

\section{Application to protostellar discs}\label{S4}
In the previous section we showed that the instability  {threshold} is determined by three parameters, $\epsilon = \Sigma_d/\Sigma_g$, $\xi = (c_d/c_g)^2$ and $\text{St} = t_s\kappa$. Here, we aim at understanding instability conditions in protostellar discs: to do so, we need to choose realistic values of these parameters. 

Firstly, the value of $\epsilon$ in protostellar discs is usually chosen $\epsilon \sim 0.01$. However, the gas disc is usually larger than the dust one, because of the radial drift: for this reason, locally, $\epsilon$ can reach higher values.

Secondly, the value of $\xi$ is more complex to determine: indeed, dust particles stirring in protoplanetary discs is due to gravitational, aerodynamical and turbulent effects. In this work, we neglect turbulent phenomena since their magnitude is smaller in these systems. A simple relation between the two parameters, taking into account only the role of drag force, can be easily found \citet{youdin}, and it reads
\begin{equation}\label{betast}
     \xi = \frac{\alpha_{SS}}{1+\text{St}},
\end{equation}
 {where $\alpha_{SS}$ is the $\alpha-$viscosity \citep{shaksun}. The relative temperature is of course related to the viscosity of the disc: indeed, the threshold below which dust particles behave the same as gas ones is given by $\text{St}<\alpha_{SS}$.}


Thirdly, the Stokes number of dust particles is essentially determined by the gas surface density and the dust particles' size. There are two main regimes of drag coupling, according to the value of the so-called Knudsen number
\begin{equation}
    \text{Kn} = \frac{9\lambda_g}{4s},
\end{equation}
where $\lambda_g$ is the gas mean free path and $s$ is the dust particles' size. Epstein regime occurs when Kn$>1$, whereas Stokes regime when Kn$<1$. Although typical protoplanetary discs are well described by the Epstein regime, self-gravitating systems are between the two regimes \citep{rice06}. It means that, for the same particles' size, the Stokes number tends to be higher in self gravitating discs\footnote{This is particularly true in the inner part of the disc, where the Knudsen number is lower since the gas density increases.}.

\subsection{Small dust particles}\label{s41}
Small dust particles are strongly coupled to the gas $(\text{St}\to0)$, thus they can be considered as a single fluid with a unique sound speed. The condition $\xi = 1$ means that $c_d = c_g$, thus drag force has no effects because of the basic state we choose: indeed, ${u}_{d0} = {u}_{g0}$, and if the dispersion velocities are the same, the response of the two fluids to the perturbations is equal $(u_{g1} =  u_{d1})$, and drag force does not act since it depends on the difference of speed. Actually, the basic velocity of gas and dust is different because of the gas pressure gradient, that is the cause of the radial drift: this issue will be discussed in the following paragraphs.

\begin{figure*}
	\includegraphics[scale=0.465]{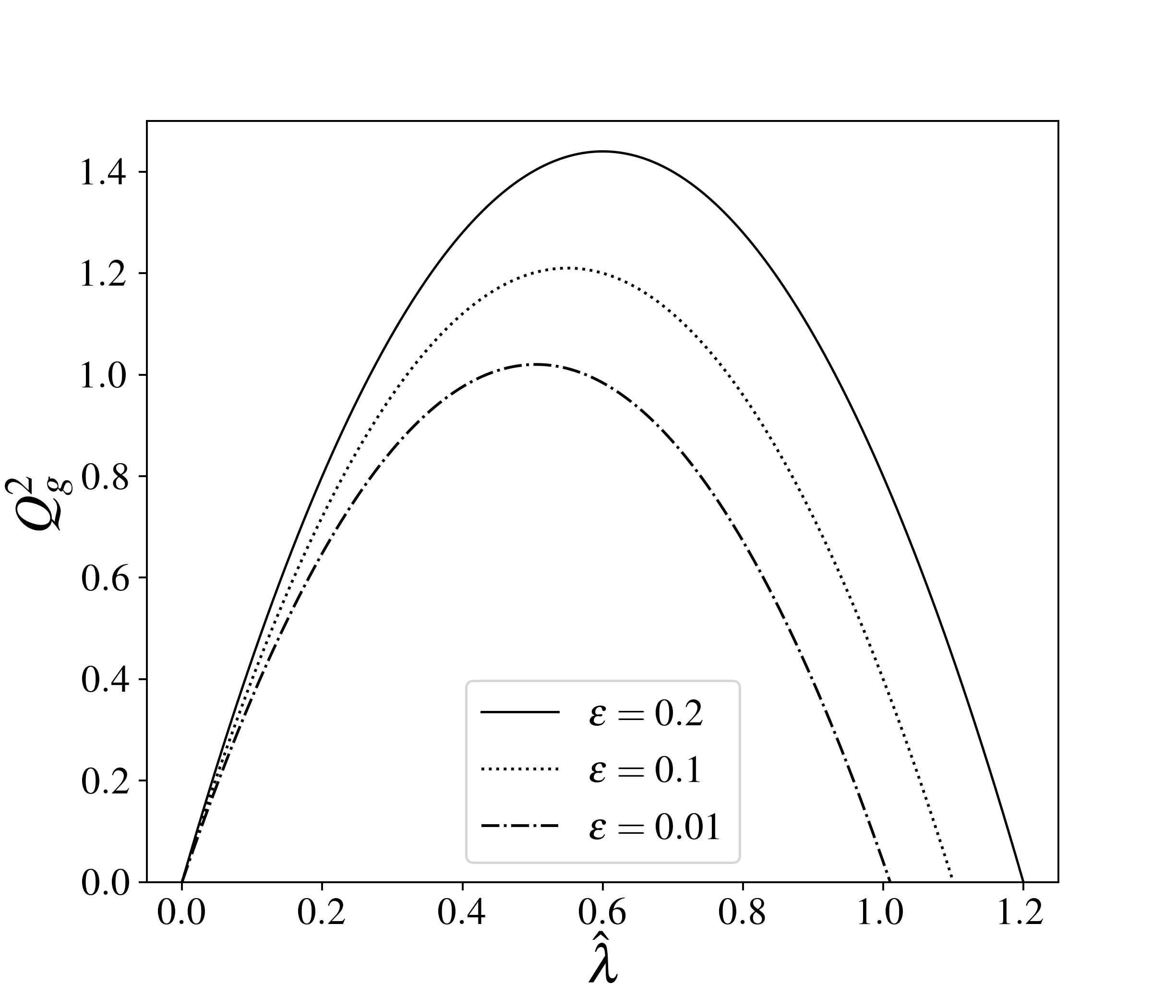}
	\includegraphics[scale=0.425]{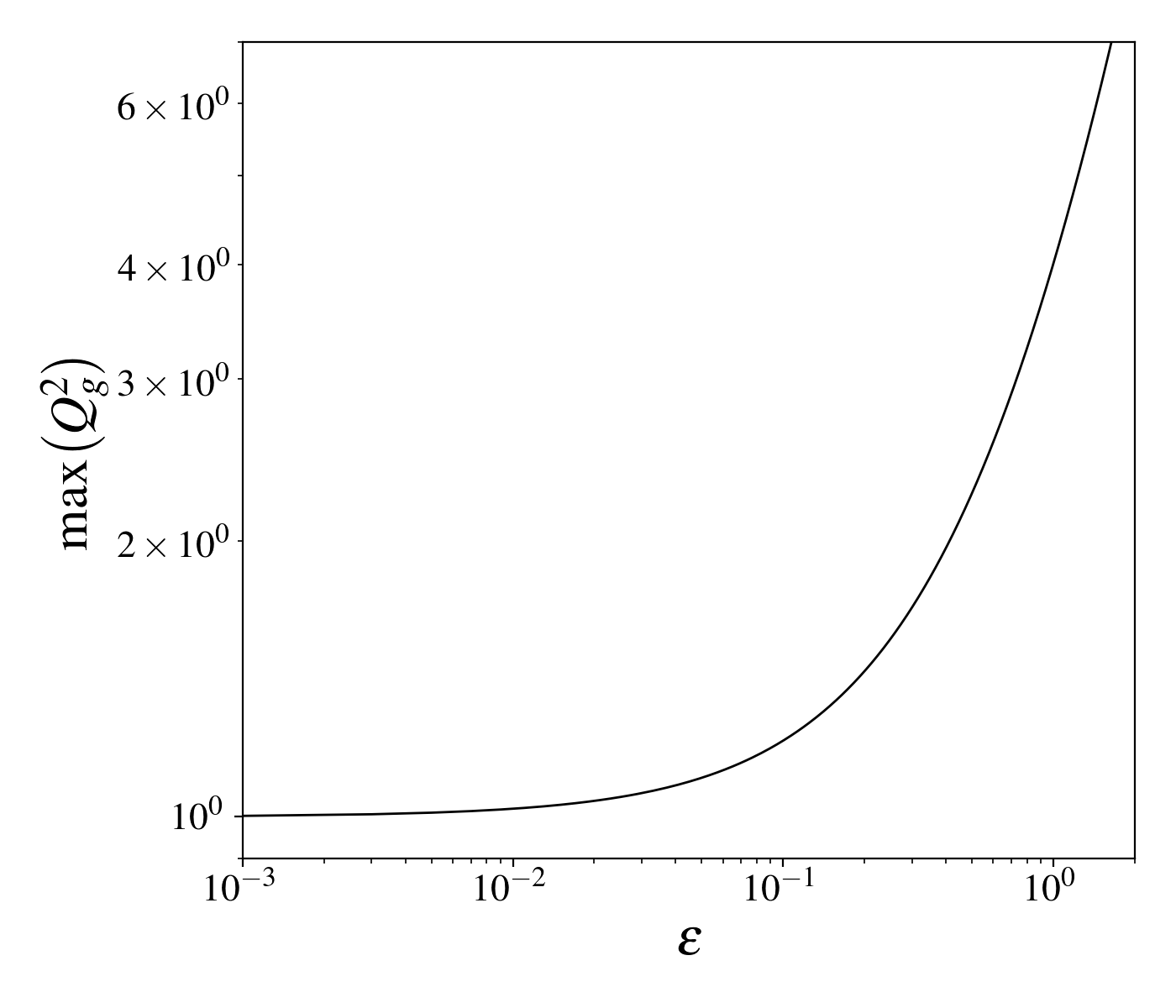}
    \caption{Stability for small dust particles. Left panel: marginal stability curve in the high drag regime with $\xi = 1$ and $\text{St}\to 0$. The parameter that controls the stability is $\epsilon$. Right panel: maximum of $Q_g^2$ as a function of the dust-to-gas ratio with $\xi = 1$ and $\text{St}\to0$. For increasing dust-to-gas ratio, the system is more unstable.}
    \label{lowst}
\end{figure*}

Hence, in this regime, the stability  {threshold} is determined only by $\epsilon$ and we recover the one fluid limit with $\Sigma_\text{tot} = \Sigma_g + \Sigma_d = \Sigma_g(1+\epsilon)$, as also shown in the context of secular GI \citep{SGI14}. In this case, the cold component has a destabilizing role, since it increases the surface density of a factor $1+\epsilon$. Figure \ref{lowst} shows the situation described so far: the left panel illustrates the marginal stability curve for different values of the dust-to-gas ratio, and the right panel shows the maximum of the curve as a function of $\epsilon$. This value represents the critical Toomre parameter, and it increases for higher dust concentration.

\subsection{Large dust particles}
Large dust particles are less coupled, and in general their sound speed is different from the gas one. As we have shown the Stokes number and $\xi$ are linked through Eq. (\ref{betast}), so we can study the instability threshold as a function of the Stokes number alone, {for different values of dust-to-gas ratio and $\alpha-$viscosity.}

\begin{figure*}
	\includegraphics[scale=0.465]{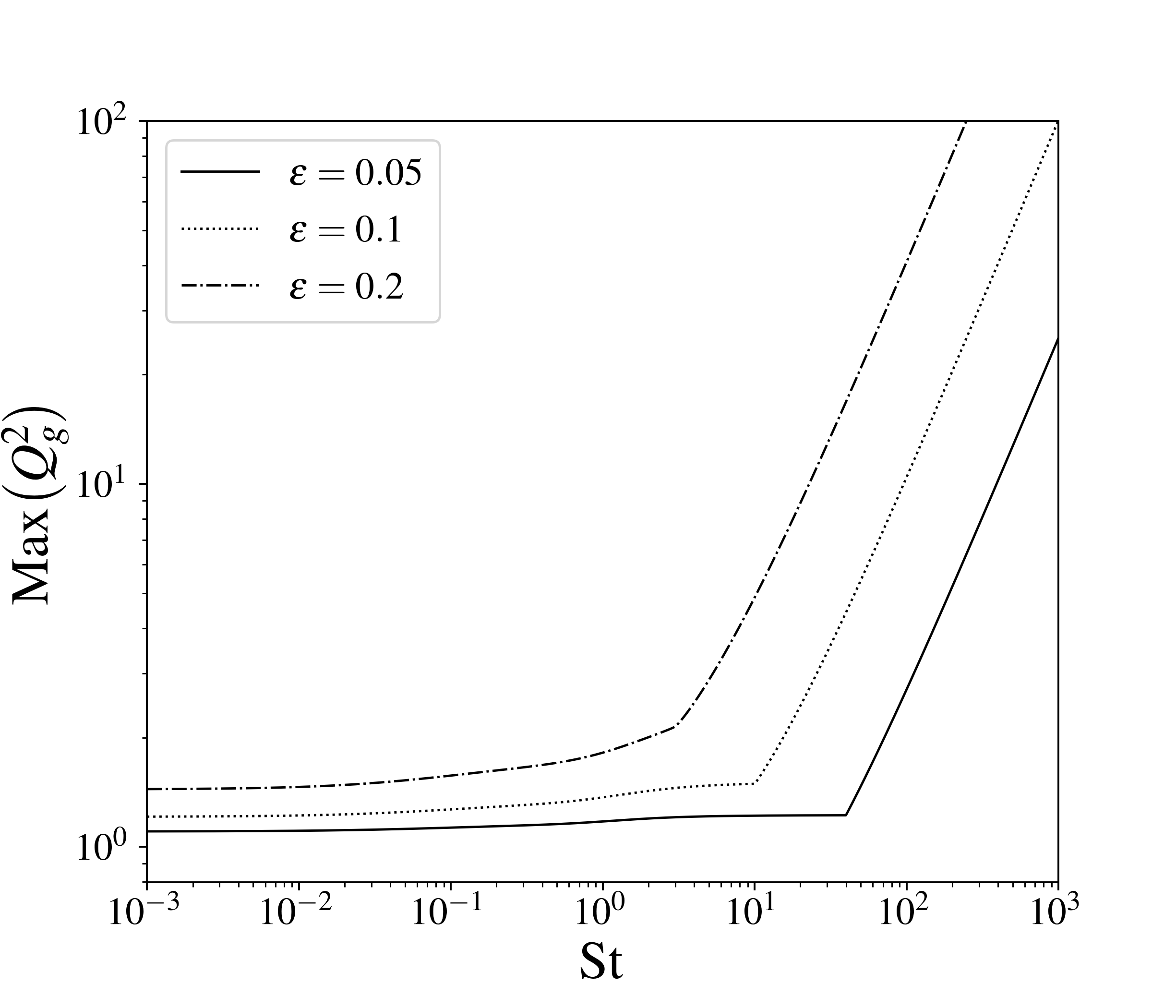}
	\includegraphics[scale=0.465]{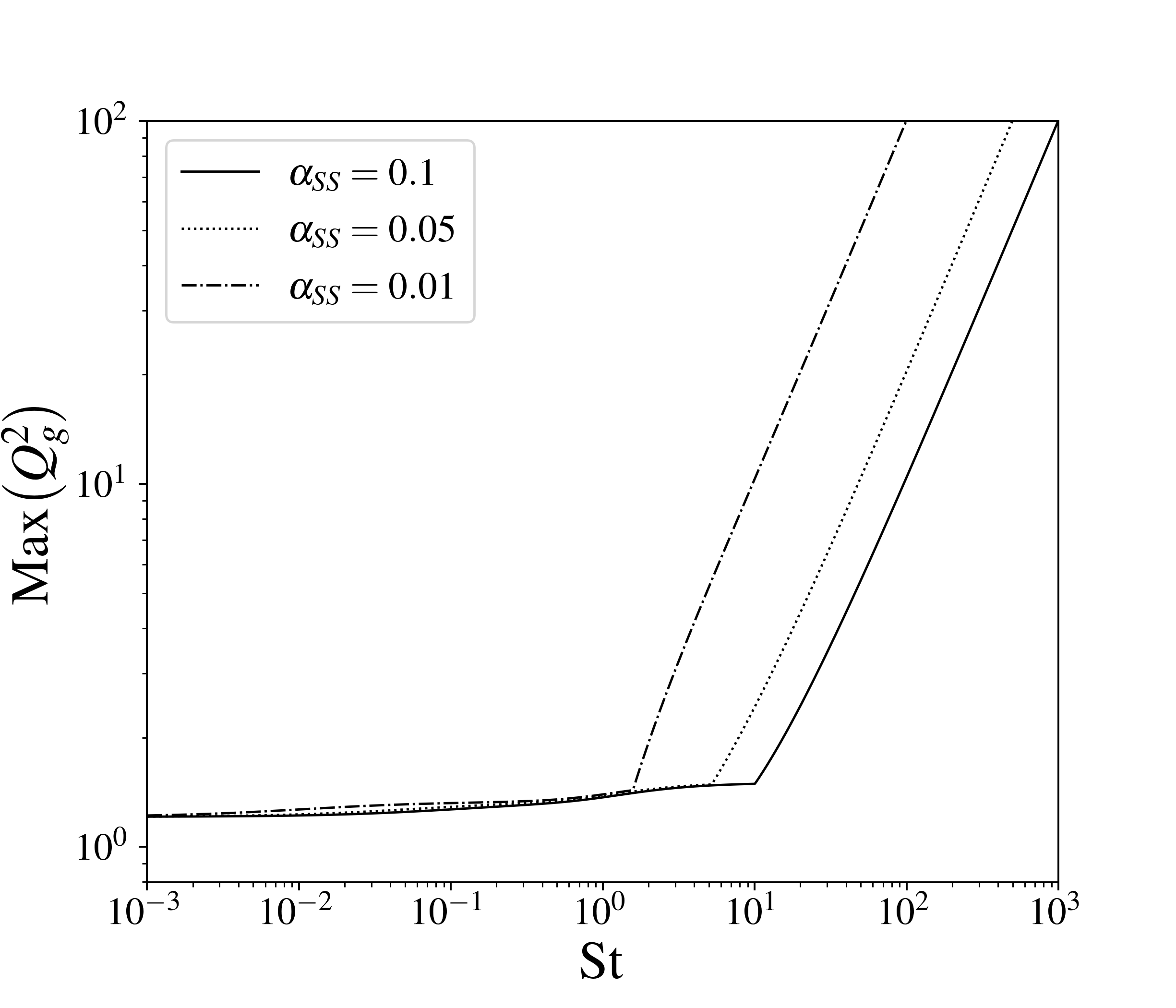}1
    \caption{ {Marginal stability curve for large dust particles, where we assumed the relation between $\xi$ and $\text{St}$ to be equation (\ref{betast}), for different values of dust to gas ratio (left panel, $\alpha_{SS}=0.1$) and the $\alpha-$viscosity (right panel, $\epsilon = 0.1$). The curve start rising for $\text{St}>1$, and it has an exponential growth when $\text{St} \gtrsim \epsilon^{-2}\alpha_{SS}$.}}
    \label{generalmsc}
\end{figure*}

Figure \ref{generalmsc} shows how the maximum of the marginal stability curve (i.e. the squared critical Toomre parameter) changes as a function of the Stokes number for different values of $\epsilon$  {and $\alpha_{SS}$. For low Stokes number ($\text{St}<\alpha_{SS}$)}, $\xi\to1$ and, we recover the previous case: the instability threshold is determined by the parameter $\epsilon$. Increasing the Stokes number, the maximum of $Q_g^2$ remains essentially constant until $\text{St}\sim1$ and then it starts rising: this happens because for $\text{St}>1$ the relation between $\xi$ and $\text{St}$ is decreasing. When $\text{St}\in[10,1000]$, depending on $\epsilon$  { and $\alpha_{SS}$}, the curve rises exponentially: this happens because there is a transition from gas-driven instability (gaseous peak) to dust-driven instability (dusty peak).  {Additionally, the lower the $\alpha-$viscosity is, the sooner the transition between gas to dust driven GI happens. In general, for a disc in gravito-turbulent regime, the value of $\alpha_{SS}$ can be relatively high $\alpha_{SS}\sim 0.05-0.1$ \citep{lodkr}, being GI an effective way to transfer angular momentum.}

It is possible to find an approximate relation between the critical Stokes number for which the instability is driven by the dust and $\epsilon$ parameter, that reads
\begin{equation}
    \text{St}_\text{crit} \simeq \epsilon^{-2}\alpha_{SS},
\end{equation}
where the last relation is obtained through a fit.  {Hence, systems with smaller dust-to-gas ratio and higher $\alpha-$viscosity show the transition at higher Stokes number, being more stable.}


\subsection{Planetesimal formation through gravitational instability}

\begin{figure*}
	\includegraphics[scale=0.465]{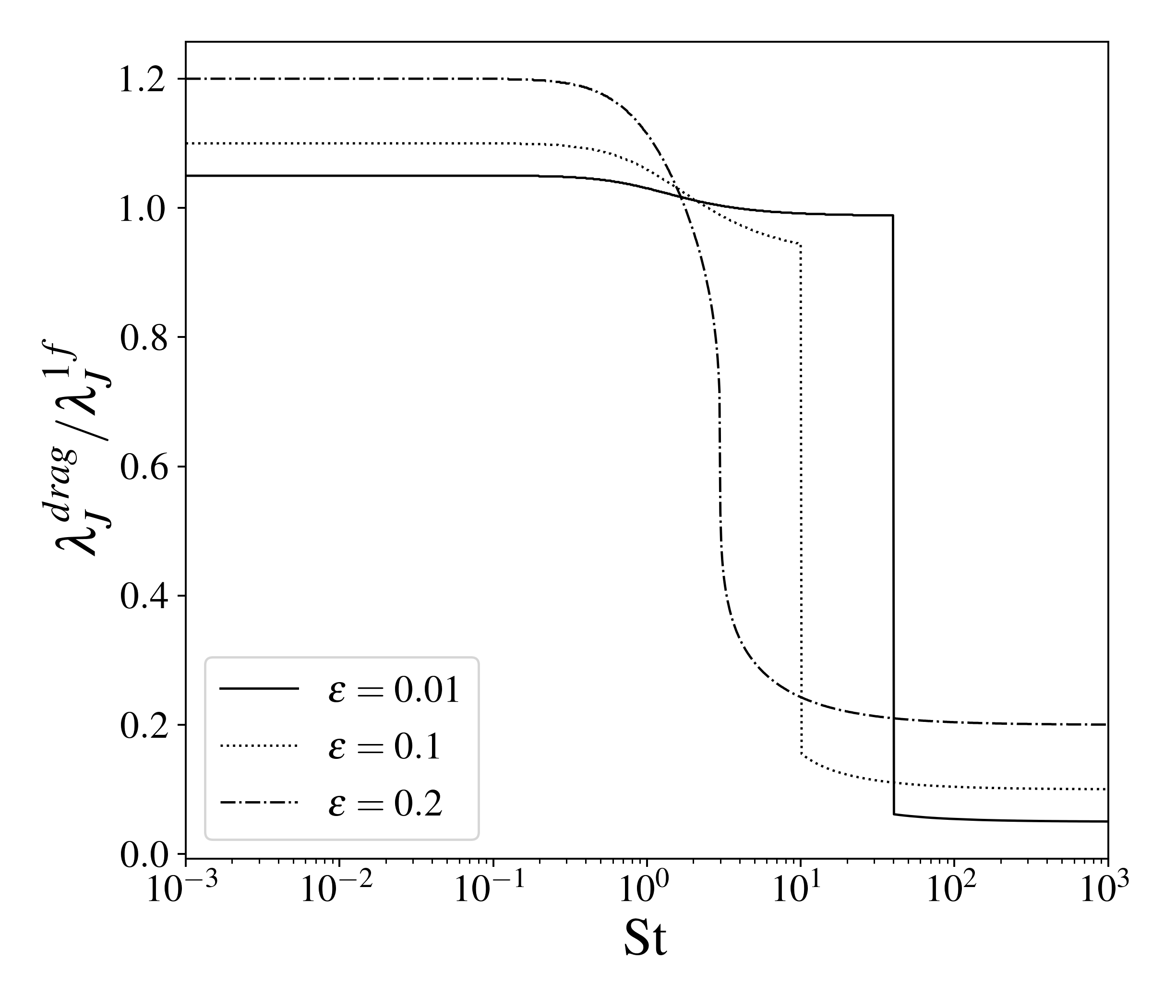}
	\includegraphics[scale=0.465]{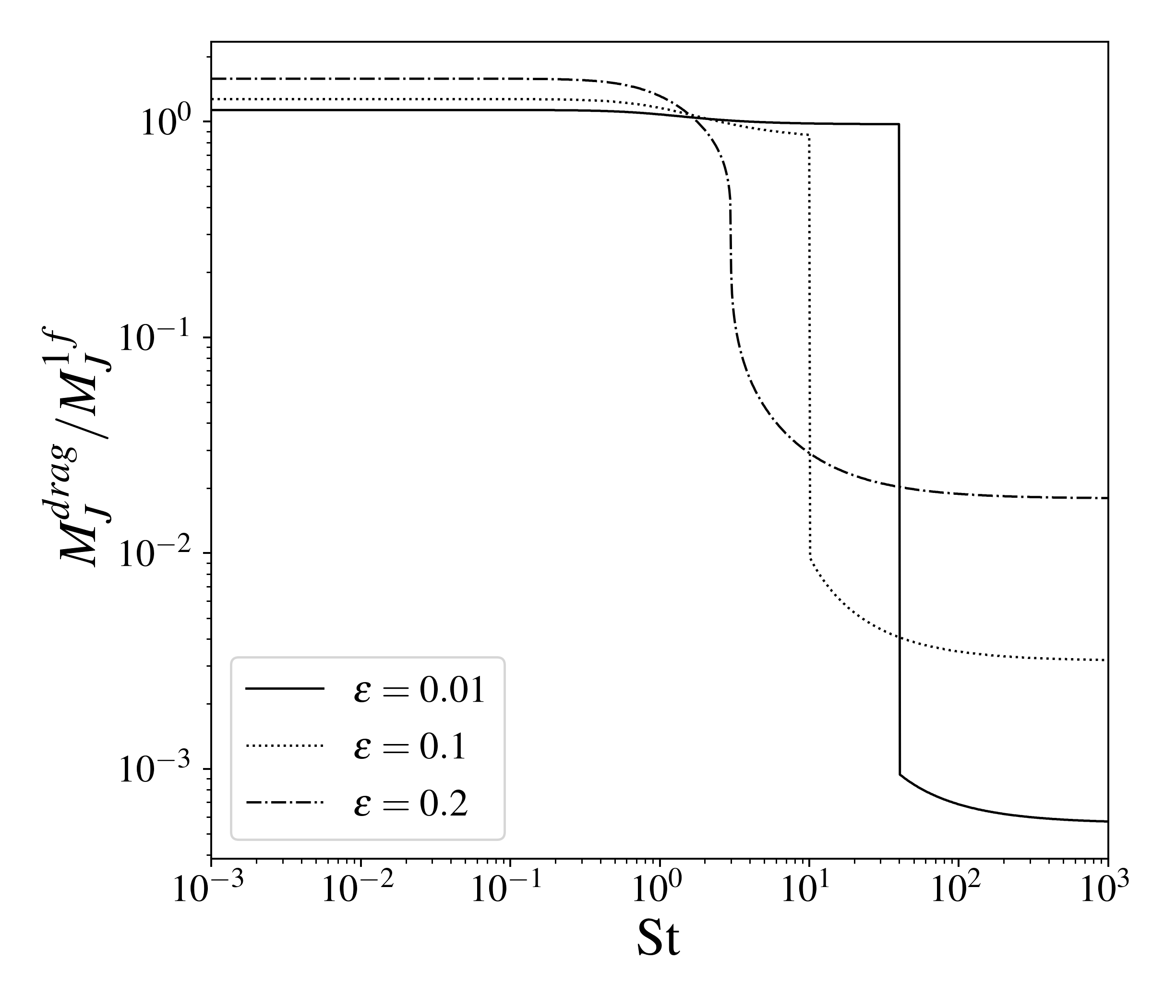}
    \caption{Jeans length (left panel) and Jeans mass (right panel) normalized to the LS one as a function of the Stokes number, for varying dust to gas ratio and $\alpha_{SS} =0.1$. For typical protoplanetary disc parameters, the Jeans mass $\mu_\text{J}\sim 1M_j$, meaning that dust-triggered instability could potentially lead to fragments of several Earth masses.}
    \label{jeans}
\end{figure*}

Now that we have obtained a general relation between $\xi$ and St, we can study the Jeans length and Jeans mass of the perturbation. The left panel of figure \ref{jeans} shows the Jeans length of our model $\lambda_J ^\text{drag}$ normalized to the one-component fluid model one $\lambda_\text{J} ^\text{1f}$, and, as expected, for $\text{St}>\epsilon^{-2}$ the value of $\lambda_\text{J}^\text{drag}$ decreases because the instability becomes dust-driven. The right panel of figure \ref{jeans} shows the Jeans mass of our model $M_\text{J}^\text{drag}$  normalized to the one fluid one $M_\text{J}^\text{1f}$. As for the Jeans wavelength, $M_\text{J}^\text{drag}$ decreases when $\text{St}>\epsilon^{-2}$, reaching values of $\sim10^{-3}$.

The classical framework\footnote{With ``classical framework'' we refer to the case of a gas-only disc, with Jeans length computed from one-component fluid model.} of gravitational instability (for a review see \citet{lodkr}) can not explain the formation of planets, since the value of the Jeans mass is too high.

Within our model, it is possible to obtain Earth-like bodies through gravitational instability when it is dust-driven: as a matter of fact, the right panel of figure \ref{jeans} shows that $M_\text{J}^\text{drag}\sim 10^{-2/-3}M_\text{J}^\text{1f}$, leading to a clump mass of the order of several Earth masses.


 {A mechanism that may allow the formation of Earth-like bodies as a consequence of Gravitational Instability (GI) was  proposed by \cite{rice06}. Concentration of solid particles in spiral arms, together with vertical settling, can lead to gravitational collapse in the solid component. Our findings are in agreement with the work \citet{rice06}, however, higher resolution is needed in order to properly assess the mass of the fragments. In addition, it is known that gas spiral arms act as dust traps, since they are pressure maxima \citep{Shi16}. Solid particles collect inside them \citep{dipierrotrap}, reaching dust concentration that can be of the order of unity. Conversely, the interaction between gas spiral arms and dust particles can excite them, imparting random motions that reduce the peak density and potential for collapse \citep{Riols20}. \cite{kicks} found that large dust particles experience gravitational scattering by the spiral arms, while \cite{boothclarke} related the level of excitation of solid particles with the aerodynamical coupling and the cooling factor. Marginally coupled solid particles are less excited by spiral arms, while, in rapidly cooled discs, the level of dust excitation is higher. \cite{baehr} confirmed that trend through 3D shearing box simulations.}

 {A possible scenario in which dust driven GI can be promoted is during stellar flybys \citep{reviewflyby}. Indeed, a flyby can rip away the external part of a protostellar disc that, because of radial drift, is low in dust \citep{flyby1}. Hence, after the interaction with the perturber, the dust to gas ratio in the inner part is significantly higher, fostering dust driven GI and, possibly, formation of planetesimals.}

\subsection{Comparison with previous works}
 {In protoplanetary discs literature, the interplay between drag force and gravitational instability has been studied in the context of ``Secular gravitational instability" \citep{SG11,SGI14,SGI20}, hereafter SGI. The general formulation of SGI is presented in \cite{SG11}. In this model, dust particles are subject to turbulent diffusion (Eq. 11a of \cite{SG11}), dust self gravity, drag force, and they are characterized by random velocities. The gas background is stable, and the role of its self gravity is negligible (i.e. $Q_g \to \infty$). The results are that there is a low frequency mode instability, that corresponds to a secular time instability, responsible for the creation of dust overdensities in a gravitationally stable gas disc ($Q_g>>1$). More refined models have been presented in the following years \citep{SGI14,SGI20}, however the crucial difference between SGI and the model we propose here is that drag force is essential for SGI to arise, while in our model the instability already exists in the limit of $\text{St}\to\infty$ \citep{jogsolomon,bertinromeo}. Moreover, our model investigates the gravitational stability of a two fluid system, and we are interested in understanding how the presence of dust destabilizes the system. In this work, the two components have different sound speed and are coupled through gravitational and drag force, without turbulent diffusion. The origin of gas and dust sound speed is different: as for the gas, it is generated by collision of particles (thermal origin), while for the dust, it is caused by stirring processes. In addition, the instability described in our model happens on a dynamical timescale, not in a secular one as for SGI. We decided not to include diffusive terms due to gas turbulence since, in the hydrodynamic equations (see appendix \ref{appa}) the gas is considered as an inviscid fluid. As we have mentioned in section 3.1, the instability threshold in the limit of $\text{St}\to0$ obtained within our model is in good agreement with what secular GI predicts, when diffusion is negligible.}


\subsection{The role of the asymmetric drift}
As we have pointed out in paragraph \ref{s41}, in this analysis we are not considering the difference of velocity in the basic state between gas and dust due to the pressure gradient. In the context of protostellar discs, this is particularly important since it is the main cause of the radial drift of solid particles.

\citet{cava} obtained the dispersion relation for a self-gravitating disc made of two components in relative motion, without any coupling between them. In a rotating self-gravitating axisymmetric fluid disc at equilibrium, the radial gravitational force is balanced by rotation, with a contribution from the pressure gradient
\begin{equation}
    \Omega^2 = \frac{1}{r\Sigma}\frac{\text{d}P}{\text{d}r} + \frac{1}{r} \frac{\text{d}\Phi}{\text{d}r},
\end{equation}
where $P$ is the pressure. Since the pressure is connected to the sound speed, two fluids with different temperature have different angular frequency ($\Omega_g\neq\Omega_d$ in protostellar case). For cool fluids (small $c$), the pressure gradient is negligible, and this is the case of dust in protostellar discs, whereas it is important for hotter fluids, i.e. the gas. \citet{cava} showed that there are two new parameters that determine the stability of the system, that are
\begin{equation}
    \delta = \frac{\Omega_d}{\Omega_g},
\end{equation}
\begin{equation}
    \eta = {m(\delta-1)}{(\Omega_g/\kappa_g)}.
\end{equation}
In the case of axisymmetric disturbances, $\eta = 0$ and so the only important parameter is $\delta$. Since the hot component moves slower than the cool one because of the pressure gradient, usually $\delta>1$. It is possible to show that in this regime the instability conditions are the same as the two fluid component model \citep{jogsolomon}, with an appropriate rescaling\footnote{ $\epsilon \to \epsilon / \delta^2$ and $\xi\to\xi/\delta^2$.  {Formally, for axisymmetric perturbations, the required rescaling indicates that a two-component model in which the presence of asymmetric drift is recognized explicitly is more stable than a model without such drift and with the same values of surface densities, temperatures, and rotation curve (because the corresponding value of the effective Q which determines marginal stability is lower for the model with the drift).}} of $\epsilon$ and $\xi$. 

 {In turn, for $m\neq0$ a two-component disk in which the presence of asymmetric drift is recognized explicitly, taken to be marginally stable with respect to m = 0 perturbations, may be unstable even for small values of $\eta$.} This fact is particularly interesting in protostellar discs: indeed, the number of spiral arms generated by gravitational instability $m$ is inversely proportional to the disc-to-star mass ratio $q$ \citep{cossins}, and thus we expect to find $m>>1$ for relatively light discs. 

\begin{figure*}
	\includegraphics[scale=0.465]{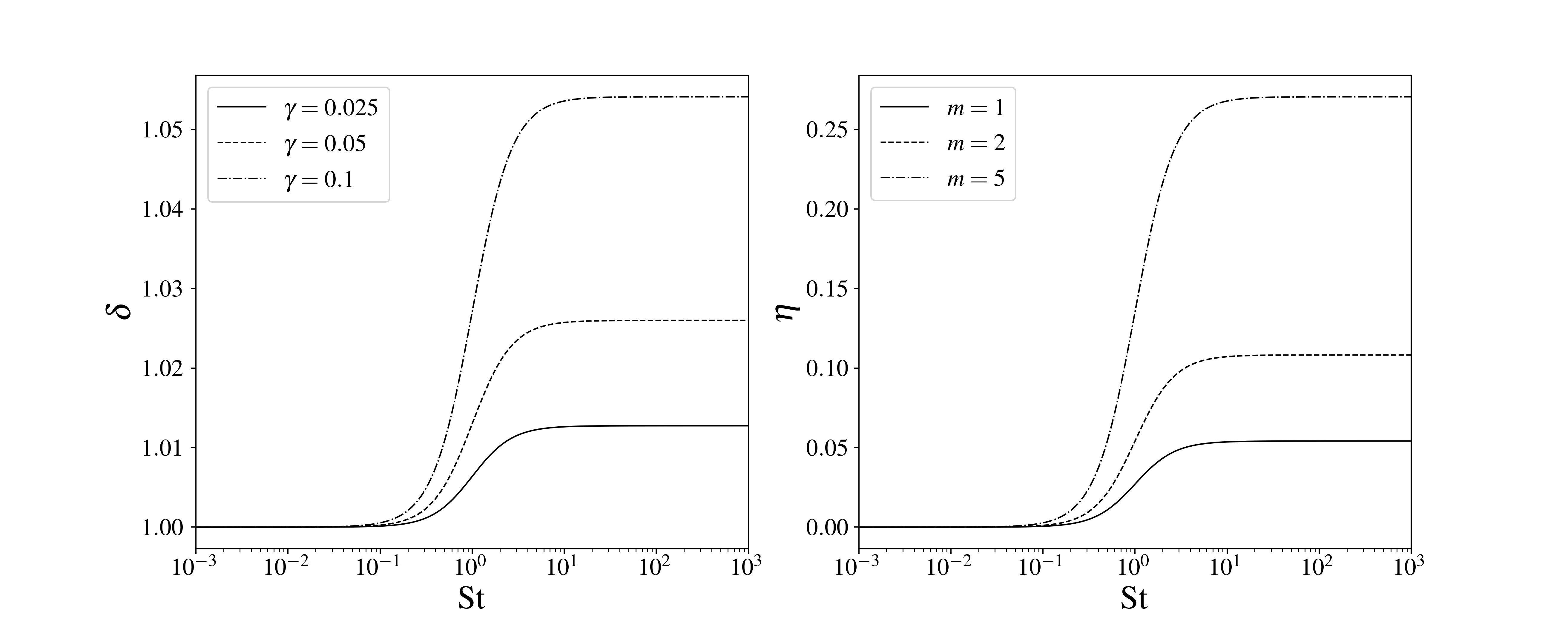}
    \caption{$\delta$ and $\eta$ parameters for a protostellar disc. Left panel: ratio between dust and gas azimuthal velocity in the steady state, i.e. $\delta$, for different values of $\gamma$, i.e. pressure gradient, as a function of the Stokes number. For St$\to0$, the dust velocity is equal to the gas one, since the two components are strongly coupled. Conversely, for St$\to\infty$, dust velocity is higher than the gaseous one, and it tends to the Keplerian speed $u_k$ and the asymmetric drift is maximum. Right panel: $\eta$ parameter as a function for the Stokes number for different values of the azimuthal wavenumber $m$, for $\gamma = 0.1$ and $\Omega_g/\kappa_g = 1$.}
    \label{asymm_drift}
\end{figure*}

In protoplanetary discs, we are able to exactly determine the asymmetric drift, and thus the value of $\delta$ and $\eta$. To do so, we neglect the dust pressure gradient, and we consider the two components to be coupled through the drag force. It is possible to show \citep{bookarmit} that, in absence of gas radial motion, the azimuthal velocities are
\begin{equation}
    u_{g0} = u_k \left(1-\gamma\right)^{1/2},
\end{equation}
\begin{equation}
    u_{d0} = \frac{\text{St}^2}{1+\text{St}^2}\left[u_k - u_{g0} \left(1-\frac{1+\text{St}^2}{\text{St}^2}\right)   \right],
\end{equation}
where $u_k = \sqrt{GM_\star/r}$, and $\gamma$ is proportional to the disc temperature, and it is a positive quantity. Hence, the strength of the asymmetric drift is connected to the Stokes number: when gas and dust are strongly coupled ($\text{St}<<1$), they move with the same velocity, and thus the asymmetric drift is zero ($\delta=1,\eta = 0$), conversely when they are uncoupled ($\text{St}>>1$), the asymmetric drift is maximum. Hence, the value of $\delta$ in protostellar discs is simply
\begin{equation}
    \delta = \frac{u_{d0}}{u_{g0}}.
\end{equation}
Figure \ref{asymm_drift} shows $\delta$ and $\eta$ parameters for a protoplanetary disc as a function of the Stokes number. Even if we take extreme values of $\delta=1.05$ and $\eta = 0.25$ and we use the dispersion relation without drag\footnote{By what we have shown in this work, if we do not take into account drag interaction, the system will always be more unstable compared to the drag case. Hence, by evaluating the instability threshold with \cite{cava}, we give an upper limit to the instability boundary.} obtained by \cite{cava}, the marginal stability curve does not change significantly. Thus, even though the asymmetric drift is crucial in protoplanetary disc evolution, since it causes the radial drift, in terms of gravitational instability it can be neglected, at a linear level.

\subsection{Non-linear evolution}

If we consider a perturbed disc, the value of $\epsilon$ and $\xi$ significantly changes because of the spiral density wave. As for the gas-to-dust ratio, it increases inside the spirals for two reasons: firstly, because the perturbation is a minimum of gravitational potential and secondly since it is a gas pressure maximum, and thus the dust experiences trapping. In addition, as pointed out by \cite{rice04}, dust growth is accelerated inside GI spirals, since its density is enhanced (the dust-to-gas ratio $\epsilon$ can reach values of the order of unity) and because of the effect of gravitational focusing. As for $\xi$, \cite{boothclarke} computed the dust dispersion velocity for gravito-turbulent discs\footnote{In this work, the authors considered a marginally unstable disc with a constant cooling rate $\beta_\text{cool} = \Omega t_\text{cool}$.} as a function of the Stokes number, and they found that the gravitational potential perturbation is effective only for $\text{St}\gtrsim1$, and in this regime $\xi\propto\text{St}^{1/2}$. The minimum $\xi$ is reached for $\text{St}\sim1$, because dust particles are forced to stay into the spiral arms by both the effect of the gravitational potential and drag force. There, the dust to gas ratio is of the order of unity, and the median dispersion velocity of dust particles is $c_d\simeq10^{-1}c_g$. 

One can think to locally apply our linear theory with the parameters of the perturbed disc, using $\epsilon = 1$, $\xi = 0.01$ and $\text{St}=1$. In this regime, the instability is dust driven and the most unstable wavelength is $\hat{\lambda}\simeq 0.2$ where the critical value of $Q_{\text{g,cr}} \simeq 6$; the Jeans mass is between 2 and 3 order of magnitude lower than the one fluid one, that corresponds approximately to $M_J\sim10M_\oplus$.

 {However, we should be cautious with these results, since we are using a linear theory to describe the non-linear evolution of the system.} To properly investigate the non-linear evolution of the system, numerical simulations of gas and dust discs are needed.

\subsection{Drag force in the context of galactic dynamics}

A natural comparison can be made in the context of galactic dynamics. In disc galaxies, drag force is connected to the phenomenon of dynamical friction \citep{Chandra}. \cite{dynfrict} evaluated this effect on a star travelling at a velocity $v_\star$ through a uniform gaseous medium with sound speed $c_g$: gaseous drag is generally more efficient when $\mathcal{M}_\star = v_\star/c_g>1$, meaning that the star motion is supersonic. This is a particularly interesting case, since gas sound speed in the Milky Way is $c_g \simeq \SI{10}{km/s}$ \citep{Fux} and the typical star velocity is $v_\star\simeq\SI{30}{km/s}$. Although drag force in galactic environments is less relevant than in protostellar ones, it would be worth to evaluate its impact on global spiral modes.

\section{Conclusions}\label{S5}
In this work, we study the dynamical role of drag force in gravitational instability, and we propose a path to form planetesimals in early protoplanetary stages. The problem of the classical Gravitational Instability scenario of planet formation \citep{bossGI} is that the Jeans mass is too large to form a planet, conversely it is an effective way to form low mass stellar companions. 

The classical GI consider the system composed of one fluid, however, protoplanetary discs are made up of two components, gas and dust. When we consider the dynamical role of the second component, GI outcome can significantly change. Indeed, the instability threshold is always higher, and the presence of the second cold can trigger instability at very short wavelengths, reducing of several order of magnitudes the Jeans length and mass.

Nevertheless, a step forward can be made: indeed, gas and dust in protoplanetary discs are aerodynamically coupled, and the role of drag force is crucial in determining their dynamical evolution. When we take into account the coupling between the two components (section \ref{S3}), gravitational instability  {threshold} is determined by three parameters, that are the relative concentration of the two fluids $\epsilon$, the relative temperature $\xi$ and the Stokes number, that measures the strength of the aerodynamical coupling. The effect of drag force in terms of gravitational instability is to connect one-component fluid model and two-component fluid model: in particular, if  drag coupling is strong, the system behaves as a one fluid, conversely, if the two components are poorly coupled, the system behaves as two-component fluid. 

We then applied this model to investigate gravitational instability in protoplanetary discs (section \ref{S4}). We first hypothesize that dust velocity dispersion is completely determined by stirring processes, so that it can be written as a function of the Stokes number. Within this hypothesis, we found that instability is dust driven when $\text{St}>\epsilon^{-2}$: hence, the Jeans mass is 3-4 order of magnitude smaller; thus, dust driven gravitational instability can be a viable way to form planetesimals in massive protostellar systems. 

In addition, we studied the role of the asymmetric drift, that in protostellar discs is significant: we quantified its effect, and we stated that it does not impact on the value of the most unstable wavelength, but only on the critical $Q_g$. Then, we discussed the non-linear evolution of the system, showing that GI spirals significantly modify the value of the stability parameters. We made a comparison between our model and numerical simulations of \cite{rice06,boothclarke} and we found good agreement. However, we should pay attention to this because our theory describes the linear behaviour of the system.

To conclude, we briefly discussed a possible application of this work in the context of galactic dynamics.


\section*{Acknowledgments}
The authors thank the anonymous referee for suggestions and comments, that significantly improved the quality of this work. This project and the authors have received funding from the European Union’s Horizon 2020 research and innovation programme under the Marie Skłodowska-Curie grant agreement N. 823823 (DUSTBUSTERS RISE project). CL acknowledges funding from Fulbright Commission through VRS scholarship. CL thanks Callum Fairbairn for fruitful discussions.

\section*{Data Availability}

The data presented in this article will be shared on reasonable request
to the corresponding author.


\bibliography{biblio}

\begin{thebibliography}{}
\makeatletter
\relax
\def\mn@urlcharsother{\let\do\@makeother \do\$\do\&\do\#\do\^\do\_\do\%\do\~}
\def\mn@doi{\begingroup\mn@urlcharsother \@ifnextchar [ {\mn@doi@}
  {\mn@doi@[]}}
\def\mn@doi@[#1]#2{\def\@tempa{#1}\ifx\@tempa\@empty \href
  {http://dx.doi.org/#2} {doi:#2}\else \href {http://dx.doi.org/#2} {#1}\fi
  \endgroup}
\def\mn@eprint#1#2{\mn@eprint@#1:#2::\@nil}
\def\mn@eprint@arXiv#1{\href {http://arxiv.org/abs/#1} {{\tt arXiv:#1}}}
\def\mn@eprint@dblp#1{\href {http://dblp.uni-trier.de/rec/bibtex/#1.xml}
  {dblp:#1}}
\def\mn@eprint@#1:#2:#3:#4\@nil{\def\@tempa {#1}\def\@tempb {#2}\def\@tempc
  {#3}\ifx \@tempc \@empty \let \@tempc \@tempb \let \@tempb \@tempa \fi \ifx
  \@tempb \@empty \def\@tempb {arXiv}\fi \@ifundefined
  {mn@eprint@\@tempb}{\@tempb:\@tempc}{\expandafter \expandafter \csname
  mn@eprint@\@tempb\endcsname \expandafter{\@tempc}}}

\bibitem[\protect\citeauthoryear{{Andrews} et~al.,}{{Andrews}
  et~al.}{2018}]{dharp1}
{Andrews} S.~M.,  et~al., 2018, \mn@doi [\apjl] {10.3847/2041-8213/aaf741},
  \href {https://ui.adsabs.harvard.edu/abs/2018ApJ...869L..41A} {869, L41}

\bibitem[\protect\citeauthoryear{{Armitage}}{{Armitage}}{2013}]{bookarmit}
{Armitage} P.~J.,  2013, {Astrophysics of Planet Formation}

\bibitem[\protect\citeauthoryear{{Baehr} \& {Zhu}}{{Baehr} \&
  {Zhu}}{2021}]{baehr}
{Baehr} H.,  {Zhu} Z.,  2021, \mn@doi [\apj] {10.3847/1538-4357/abddb3}, \href
  {https://ui.adsabs.harvard.edu/abs/2021ApJ...909..135B} {909, 135}

\bibitem[\protect\citeauthoryear{{Bertin} \& {Cava}}{{Bertin} \&
  {Cava}}{2006}]{cava}
{Bertin} G.,  {Cava} A.,  2006, \mn@doi [\aap] {10.1051/0004-6361:20065012},
  \href {https://ui.adsabs.harvard.edu/abs/2006A&A...459..333B} {459, 333}

\bibitem[\protect\citeauthoryear{{Bertin} \& {Lodato}}{{Bertin} \&
  {Lodato}}{1999}]{bertinlod}
{Bertin} G.,  {Lodato} G.,  1999, \aap, \href
  {https://ui.adsabs.harvard.edu/abs/1999A&A...350..694B} {350, 694}

\bibitem[\protect\citeauthoryear{{Bertin} \& {Romeo}}{{Bertin} \&
  {Romeo}}{1988}]{bertinromeo}
{Bertin} G.,  {Romeo} A.~B.,  1988, \aap, \href
  {https://ui.adsabs.harvard.edu/abs/1988A&A...195..105B} {195, 105}

\bibitem[\protect\citeauthoryear{{Bollati}, {Lodato}, {Price}  \&
  {Pinte}}{{Bollati} et~al.}{2021}]{kinkbollati}
{Bollati} F.,  {Lodato} G.,  {Price} D.~J.,   {Pinte} C.,  2021, \mn@doi
  [\mnras] {10.1093/mnras/stab1145}, \href
  {https://ui.adsabs.harvard.edu/abs/2021MNRAS.504.5444B} {504, 5444}

\bibitem[\protect\citeauthoryear{{Booth} \& {Clarke}}{{Booth} \&
  {Clarke}}{2016}]{boothclarke}
{Booth} R.~A.,  {Clarke} C.~J.,  2016, \mn@doi [\mnras] {10.1093/mnras/stw488},
  \href {https://ui.adsabs.harvard.edu/abs/2016MNRAS.458.2676B} {458, 2676}

\bibitem[\protect\citeauthoryear{{Boss}}{{Boss}}{1997}]{bossGI}
{Boss} A.~P.,  1997, \mn@doi [Science] {10.1126/science.276.5320.1836}, \href
  {https://ui.adsabs.harvard.edu/abs/1997Sci...276.1836B} {276, 1836}

\bibitem[\protect\citeauthoryear{{Chandrasekhar}}{{Chandrasekhar}}{1943}]{Chandra}
{Chandrasekhar} S.,  1943, \mn@doi [\apj] {10.1086/144517}, \href
  {https://ui.adsabs.harvard.edu/abs/1943ApJ....97..255C} {97, 255}

\bibitem[\protect\citeauthoryear{{Cossins}, {Lodato}  \& {Clarke}}{{Cossins}
  et~al.}{2009}]{cossins}
{Cossins} P.,  {Lodato} G.,   {Clarke} C.~J.,  2009, \mn@doi [\mnras]
  {10.1111/j.1365-2966.2008.14275.x}, \href
  {https://ui.adsabs.harvard.edu/abs/2009MNRAS.393.1157C} {393, 1157}

\bibitem[\protect\citeauthoryear{{Crida}, {Morbidelli}  \& {Masset}}{{Crida}
  et~al.}{2006}]{morph1}
{Crida} A.,  {Morbidelli} A.,   {Masset} F.,  2006, \mn@doi [\icarus]
  {10.1016/j.icarus.2005.10.007}, \href
  {https://ui.adsabs.harvard.edu/abs/2006Icar..181..587C} {181, 587}

\bibitem[\protect\citeauthoryear{{Cuello} et~al.,}{{Cuello}
  et~al.}{2019}]{flyby1}
{Cuello} N.,  et~al., 2019, \mn@doi [\mnras] {10.1093/mnras/sty3325}, \href
  {https://ui.adsabs.harvard.edu/abs/2019MNRAS.483.4114C} {483, 4114}

\bibitem[\protect\citeauthoryear{{Cuello}, {M{\'e}nard}  \& {Price}}{{Cuello}
  et~al.}{2022}]{reviewflyby}
{Cuello} N.,  {M{\'e}nard} F.,   {Price} D.~J.,  2022, arXiv e-prints, \href
  {https://ui.adsabs.harvard.edu/abs/2022arXiv220709752C} {p. arXiv:2207.09752}

\bibitem[\protect\citeauthoryear{{Curone} et~al.,}{{Curone}
  et~al.}{2022}]{pietro}
{Curone} P.,  et~al., 2022, \mn@doi [\aap] {10.1051/0004-6361/202142748}, \href
  {https://ui.adsabs.harvard.edu/abs/2022A&A...665A..25C} {665, A25}

\bibitem[\protect\citeauthoryear{{Dipierro}, {Pinilla}, {Lodato}  \&
  {Testi}}{{Dipierro} et~al.}{2015}]{dipierrotrap}
{Dipierro} G.,  {Pinilla} P.,  {Lodato} G.,   {Testi} L.,  2015, \mn@doi
  [\mnras] {10.1093/mnras/stv970}, \href
  {https://ui.adsabs.harvard.edu/abs/2015MNRAS.451..974D} {451, 974}

\bibitem[\protect\citeauthoryear{{Draine}}{{Draine}}{2003}]{dusttogas}
{Draine} B.~T.,  2003, \mn@doi [\araa]
  {10.1146/annurev.astro.41.011802.094840}, \href
  {https://ui.adsabs.harvard.edu/abs/2003ARA&A..41..241D} {41, 241}

\bibitem[\protect\citeauthoryear{{Dr{\k{a}}{\.z}kowska}, {Windmark}  \&
  {Dullemond}}{{Dr{\k{a}}{\.z}kowska} et~al.}{2014}]{dustgrowth}
{Dr{\k{a}}{\.z}kowska} J.,  {Windmark} F.,   {Dullemond} C.~P.,  2014, \mn@doi
  [\aap] {10.1051/0004-6361/201423708}, \href
  {https://ui.adsabs.harvard.edu/abs/2014A&A...567A..38D} {567, A38}

\bibitem[\protect\citeauthoryear{{Facchini}, {Teague}, {Bae}, {Benisty},
  {Keppler}  \& {Isella}}{{Facchini} et~al.}{2021}]{facchiniPDS}
{Facchini} S.,  {Teague} R.,  {Bae} J.,  {Benisty} M.,  {Keppler} M.,
  {Isella} A.,  2021, arXiv e-prints, \href
  {https://ui.adsabs.harvard.edu/abs/2021arXiv210108369F} {p. arXiv:2101.08369}

\bibitem[\protect\citeauthoryear{{Fedele}, {Toci}, {Maud}  \&
  {Lodato}}{{Fedele} et~al.}{2021}]{davideHD}
{Fedele} D.,  {Toci} C.,  {Maud} L.,   {Lodato} G.,  2021, \mn@doi [\aap]
  {10.1051/0004-6361/202141278}, \href
  {https://ui.adsabs.harvard.edu/abs/2021A&A...651A..90F} {651, A90}

\bibitem[\protect\citeauthoryear{{Fux}}{{Fux}}{1999}]{Fux}
{Fux} R.,  1999, \aap, \href
  {https://ui.adsabs.harvard.edu/abs/1999A&A...345..787F} {345, 787}

\bibitem[\protect\citeauthoryear{{Goldreich} \& {Ward}}{{Goldreich} \&
  {Ward}}{1973}]{CA2}
{Goldreich} P.,  {Ward} W.~R.,  1973, \mn@doi [\apj] {10.1086/152291}, \href
  {https://ui.adsabs.harvard.edu/abs/1973ApJ...183.1051G} {183, 1051}

\bibitem[\protect\citeauthoryear{{Haisch}, {Lada}  \& {Lada}}{{Haisch}
  et~al.}{2001}]{NOCA1}
{Haisch} Karl~E. J.,  {Lada} E.~A.,   {Lada} C.~J.,  2001, \mn@doi [\apjl]
  {10.1086/320685}, \href
  {https://ui.adsabs.harvard.edu/abs/2001ApJ...553L.153H} {553, L153}

\bibitem[\protect\citeauthoryear{{Hall} et~al.,}{{Hall}
  et~al.}{2020}]{hallwigg}
{Hall} C.,  et~al., 2020, \mn@doi [\apj] {10.3847/1538-4357/abac17}, \href
  {https://ui.adsabs.harvard.edu/abs/2020ApJ...904..148H} {904, 148}

\bibitem[\protect\citeauthoryear{{Huang} et~al.,}{{Huang}
  et~al.}{2018}]{dsharp2}
{Huang} J.,  et~al., 2018, \mn@doi [\apjl] {10.3847/2041-8213/aaf740}, \href
  {https://ui.adsabs.harvard.edu/abs/2018ApJ...869L..42H} {869, L42}

\bibitem[\protect\citeauthoryear{{Jog} \& {Solomon}}{{Jog} \&
  {Solomon}}{1984}]{jogsolomon}
{Jog} C.~J.,  {Solomon} P.~M.,  1984, \mn@doi [\apj] {10.1086/161597}, \href
  {https://ui.adsabs.harvard.edu/abs/1984ApJ...276..114J} {276, 114}

\bibitem[\protect\citeauthoryear{{Johansen} \& {Youdin}}{{Johansen} \&
  {Youdin}}{2007}]{stream3}
{Johansen} A.,  {Youdin} A.,  2007, \mn@doi [\apj] {10.1086/516730}, \href
  {https://ui.adsabs.harvard.edu/abs/2007ApJ...662..627J} {662, 627}

\bibitem[\protect\citeauthoryear{{Kato}}{{Kato}}{1972}]{kato72}
{Kato} S.,  1972, \pasj, \href
  {https://ui.adsabs.harvard.edu/abs/1972PASJ...24...61K} {24, 61}

\bibitem[\protect\citeauthoryear{{Keppler} et~al.,}{{Keppler}
  et~al.}{2018}]{kepplerPDS}
{Keppler} M.,  et~al., 2018, \mn@doi [\aap] {10.1051/0004-6361/201832957},
  \href {https://ui.adsabs.harvard.edu/abs/2018A&A...617A..44K} {617, A44}

\bibitem[\protect\citeauthoryear{{Keppler} et~al.,}{{Keppler}
  et~al.}{2019}]{kepplerPDS19}
{Keppler} M.,  et~al., 2019, \mn@doi [\aap] {10.1051/0004-6361/201935034},
  \href {https://ui.adsabs.harvard.edu/abs/2019A&A...625A.118K} {625, A118}

\bibitem[\protect\citeauthoryear{{Kokubo} \& {Ida}}{{Kokubo} \&
  {Ida}}{1998}]{kokubo}
{Kokubo} E.,  {Ida} S.,  1998, \mn@doi [\icarus] {10.1006/icar.1997.5840},
  \href {https://ui.adsabs.harvard.edu/abs/1998Icar..131..171K} {131, 171}

\bibitem[\protect\citeauthoryear{{Kratter} \& {Lodato}}{{Kratter} \&
  {Lodato}}{2016}]{lodkr}
{Kratter} K.,  {Lodato} G.,  2016, \mn@doi [\araa]
  {10.1146/annurev-astro-081915-023307}, \href
  {https://ui.adsabs.harvard.edu/abs/2016ARA&A..54..271K} {54, 271}

\bibitem[\protect\citeauthoryear{{Kratter} \& {Matzner}}{{Kratter} \&
  {Matzner}}{2006}]{kratter1}
{Kratter} K.~M.,  {Matzner} C.~D.,  2006, \mn@doi [\mnras]
  {10.1111/j.1365-2966.2006.11103.x}, \href
  {https://ui.adsabs.harvard.edu/abs/2006MNRAS.373.1563K} {373, 1563}

\bibitem[\protect\citeauthoryear{{Kratter}, {Matzner}, {Krumholz}  \&
  {Klein}}{{Kratter} et~al.}{2010}]{kratter2}
{Kratter} K.~M.,  {Matzner} C.~D.,  {Krumholz} M.~R.,   {Klein} R.~I.,  2010,
  \mn@doi [\apj] {10.1088/0004-637X/708/2/1585}, \href
  {https://ui.adsabs.harvard.edu/abs/2010ApJ...708.1585K} {708, 1585}

\bibitem[\protect\citeauthoryear{{Lin} \& {Shu}}{{Lin} \& {Shu}}{1964}]{Linshu}
{Lin} C.~C.,  {Shu} F.~H.,  1964, \mn@doi [\apj] {10.1086/147955}, \href
  {https://ui.adsabs.harvard.edu/abs/1964ApJ...140..646L} {140, 646}

\bibitem[\protect\citeauthoryear{{Lodato} et~al.,}{{Lodato}
  et~al.}{2019}]{morph2}
{Lodato} G.,  et~al., 2019, \mn@doi [\mnras] {10.1093/mnras/stz913}, \href
  {https://ui.adsabs.harvard.edu/abs/2019MNRAS.486..453L} {486, 453}

\bibitem[\protect\citeauthoryear{{Longarini}, {Lodato}, {Toci}, {Veronesi},
  {Hall}, {Dong}  \& {Patrick Terry}}{{Longarini} et~al.}{2021}]{longwig}
{Longarini} C.,  {Lodato} G.,  {Toci} C.,  {Veronesi} B.,  {Hall} C.,  {Dong}
  R.,   {Patrick Terry} J.,  2021, \mn@doi [\apjl] {10.3847/2041-8213/ac2df6},
  \href {https://ui.adsabs.harvard.edu/abs/2021ApJ...920L..41L} {920, L41}

\bibitem[\protect\citeauthoryear{{Ostriker}}{{Ostriker}}{1999}]{dynfrict}
{Ostriker} E.~C.,  1999, \mn@doi [\apj] {10.1086/306858}, \href
  {https://ui.adsabs.harvard.edu/abs/1999ApJ...513..252O} {513, 252}

\bibitem[\protect\citeauthoryear{{Paneque-Carre{\~n}o}
  et~al.,}{{Paneque-Carre{\~n}o} et~al.}{2021}]{paneque}
{Paneque-Carre{\~n}o} T.,  et~al., 2021, \mn@doi [\apj]
  {10.3847/1538-4357/abf243}, \href
  {https://ui.adsabs.harvard.edu/abs/2021ApJ...914...88P} {914, 88}

\bibitem[\protect\citeauthoryear{{P{\'e}rez} et~al.,}{{P{\'e}rez}
  et~al.}{2016}]{perezelias}
{P{\'e}rez} L.~M.,  et~al., 2016, \mn@doi [Science] {10.1126/science.aaf8296},
  \href {https://ui.adsabs.harvard.edu/abs/2016Sci...353.1519P} {353, 1519}

\bibitem[\protect\citeauthoryear{{Pinilla} et~al.,}{{Pinilla}
  et~al.}{2021}]{CIDA1}
{Pinilla} P.,  et~al., 2021, \mn@doi [\aap] {10.1051/0004-6361/202140371},
  \href {https://ui.adsabs.harvard.edu/abs/2021A&A...649A.122P} {649, A122}

\bibitem[\protect\citeauthoryear{{Pinte} et~al.,}{{Pinte}
  et~al.}{2019}]{kinkpinte}
{Pinte} C.,  et~al., 2019, \mn@doi [Nature Astronomy]
  {10.1038/s41550-019-0852-6}, \href
  {https://ui.adsabs.harvard.edu/abs/2019NatAs...3.1109P} {3, 1109}

\bibitem[\protect\citeauthoryear{{Rice}, {Lodato}, {Pringle}, {Armitage}  \&
  {Bonnell}}{{Rice} et~al.}{2004}]{rice04}
{Rice} W.~K.~M.,  {Lodato} G.,  {Pringle} J.~E.,  {Armitage} P.~J.,   {Bonnell}
  I.~A.,  2004, \mn@doi [\mnras] {10.1111/j.1365-2966.2004.08339.x}, \href
  {https://ui.adsabs.harvard.edu/abs/2004MNRAS.355..543R} {355, 543}

\bibitem[\protect\citeauthoryear{{Rice}, {Lodato}, {Pringle}, {Armitage}  \&
  {Bonnell}}{{Rice} et~al.}{2006}]{rice06}
{Rice} W.~K.~M.,  {Lodato} G.,  {Pringle} J.~E.,  {Armitage} P.~J.,   {Bonnell}
  I.~A.,  2006, \mn@doi [\mnras] {10.1111/j.1745-3933.2006.00215.x}, \href
  {https://ui.adsabs.harvard.edu/abs/2006MNRAS.372L...9R} {372, L9}

\bibitem[\protect\citeauthoryear{{Riols}, {Roux}, {Latter}  \& {Lesur}}{{Riols}
  et~al.}{2020}]{Riols20}
{Riols} A.,  {Roux} B.,  {Latter} H.,   {Lesur} G.,  2020, \mn@doi [\mnras]
  {10.1093/mnras/staa567}, \href
  {https://ui.adsabs.harvard.edu/abs/2020MNRAS.493.4631R} {493, 4631}

\bibitem[\protect\citeauthoryear{{Safronov}}{{Safronov}}{1960}]{safr60}
{Safronov} V.,  1960, Annales d'Astrophysique, \href
  {https://ui.adsabs.harvard.edu/abs/1960AnAp...23..979S} {23, 979}

\bibitem[\protect\citeauthoryear{{Safronov}}{{Safronov}}{1969}]{CA1}
{Safronov} V.~S.,  1969, {Evoliutsiia doplanetnogo oblaka.}

\bibitem[\protect\citeauthoryear{{Scardoni}, {Booth}  \& {Clarke}}{{Scardoni}
  et~al.}{2021}]{scardoni}
{Scardoni} C.~E.,  {Booth} R.~A.,   {Clarke} C.~J.,  2021, \mn@doi [\mnras]
  {10.1093/mnras/stab854}, \href
  {https://ui.adsabs.harvard.edu/abs/2021MNRAS.504.1495S} {504, 1495}

\bibitem[\protect\citeauthoryear{{Segura-Cox} et~al.,}{{Segura-Cox}
  et~al.}{2020}]{cox}
{Segura-Cox} D.~M.,  et~al., 2020, \mn@doi [\nat] {10.1038/s41586-020-2779-6},
  \href {https://ui.adsabs.harvard.edu/abs/2020Natur.586..228S} {586, 228}

\bibitem[\protect\citeauthoryear{{Shakura}, {Sunyaev}  \&
  {Zilitinkevich}}{{Shakura} et~al.}{1978}]{shaksun}
{Shakura} N.~I.,  {Sunyaev} R.~A.,   {Zilitinkevich} S.~S.,  1978, \aap, \href
  {https://ui.adsabs.harvard.edu/abs/1978A&A....62..179S} {62, 179}

\bibitem[\protect\citeauthoryear{{Shi}, {Zhu}, {Stone}  \& {Chiang}}{{Shi}
  et~al.}{2016}]{Shi16}
{Shi} J.-M.,  {Zhu} Z.,  {Stone} J.~M.,   {Chiang} E.,  2016, \mn@doi [\mnras]
  {10.1093/mnras/stw692}, \href
  {https://ui.adsabs.harvard.edu/abs/2016MNRAS.459..982S} {459, 982}

\bibitem[\protect\citeauthoryear{{Stevenson}}{{Stevenson}}{1982}]{coreinst}
{Stevenson} D.~J.,  1982, \mn@doi [\planss] {10.1016/0032-0633(82)90108-8},
  \href {https://ui.adsabs.harvard.edu/abs/1982P&SS...30..755S} {30, 755}

\bibitem[\protect\citeauthoryear{{Takahashi} \& {Inutsuka}}{{Takahashi} \&
  {Inutsuka}}{2014}]{SGI14}
{Takahashi} S.~Z.,  {Inutsuka} S.-i.,  2014, \mn@doi [\apj]
  {10.1088/0004-637X/794/1/55}, \href
  {https://ui.adsabs.harvard.edu/abs/2014ApJ...794...55T} {794, 55}

\bibitem[\protect\citeauthoryear{{Tazzari}, {Clarke}, {Testi}, {Williams},
  {Facchini}, {Manara}, {Natta}  \& {Rosotti}}{{Tazzari} et~al.}{2021a}]{tazz1}
{Tazzari} M.,  {Clarke} C.~J.,  {Testi} L.,  {Williams} J.~P.,  {Facchini} S.,
  {Manara} C.~F.,  {Natta} A.,   {Rosotti} G.,  2021a, \mn@doi [\mnras]
  {10.1093/mnras/stab1808}, \href
  {https://ui.adsabs.harvard.edu/abs/2021MNRAS.506.2804T} {506, 2804}

\bibitem[\protect\citeauthoryear{{Tazzari} et~al.,}{{Tazzari}
  et~al.}{2021b}]{tazz2}
{Tazzari} M.,  et~al., 2021b, \mn@doi [\mnras] {10.1093/mnras/stab1912}, \href
  {https://ui.adsabs.harvard.edu/abs/2021MNRAS.506.5117T} {506, 5117}

\bibitem[\protect\citeauthoryear{{Terry}, {Hall}, {Longarini}, {Lodato},
  {Toci}, {Veronesi}, {Paneque-Carre{\~n}o}  \& {Pinte}}{{Terry}
  et~al.}{2021}]{terrywig}
{Terry} J.~P.,  {Hall} C.,  {Longarini} C.,  {Lodato} G.,  {Toci} C.,
  {Veronesi} B.,  {Paneque-Carre{\~n}o} T.,   {Pinte} C.,  2021, \mn@doi
  [\mnras] {10.1093/mnras/stab3513}, \href
  {https://ui.adsabs.harvard.edu/abs/2021MNRAS.tmp.3178T} {}

\bibitem[\protect\citeauthoryear{{Toci}, {Lodato}, {Christiaens}, {Fedele},
  {Pinte}, {Price}  \& {Testi}}{{Toci} et~al.}{2020a}]{claPDS}
{Toci} C.,  {Lodato} G.,  {Christiaens} V.,  {Fedele} D.,  {Pinte} C.,  {Price}
  D.~J.,   {Testi} L.,  2020a, \mn@doi [\mnras] {10.1093/mnras/staa2933}, \href
  {https://ui.adsabs.harvard.edu/abs/2020MNRAS.499.2015T} {499, 2015}

\bibitem[\protect\citeauthoryear{{Toci}, {Lodato}, {Fedele}, {Testi}  \&
  {Pinte}}{{Toci} et~al.}{2020b}]{claHD}
{Toci} C.,  {Lodato} G.,  {Fedele} D.,  {Testi} L.,   {Pinte} C.,  2020b,
  \mn@doi [\apjl] {10.3847/2041-8213/ab5c87}, \href
  {https://ui.adsabs.harvard.edu/abs/2020ApJ...888L...4T} {888, L4}

\bibitem[\protect\citeauthoryear{{Tominaga}, {Takahashi}  \&
  {Inutsuka}}{{Tominaga} et~al.}{2020}]{SGI20}
{Tominaga} R.~T.,  {Takahashi} S.~Z.,   {Inutsuka} S.-i.,  2020, \mn@doi [\apj]
  {10.3847/1538-4357/abad36}, \href
  {https://ui.adsabs.harvard.edu/abs/2020ApJ...900..182T} {900, 182}

\bibitem[\protect\citeauthoryear{{Toomre}}{{Toomre}}{1964}]{toomre64}
{Toomre} A.,  1964, \mn@doi [\apj] {10.1086/147861}, \href
  {https://ui.adsabs.harvard.edu/abs/1964ApJ...139.1217T} {139, 1217}

\bibitem[\protect\citeauthoryear{{Veronesi} et~al.,}{{Veronesi}
  et~al.}{2020}]{benniDS}
{Veronesi} B.,  et~al., 2020, \mn@doi [\mnras] {10.1093/mnras/staa1278}, \href
  {https://ui.adsabs.harvard.edu/abs/2020MNRAS.495.1913V} {495, 1913}

\bibitem[\protect\citeauthoryear{{Veronesi}, {Paneque-Carre{\~n}o}, {Lodato},
  {Testi}, {P{\'e}rez}, {Bertin}  \& {Hall}}{{Veronesi}
  et~al.}{2021}]{bennielias}
{Veronesi} B.,  {Paneque-Carre{\~n}o} T.,  {Lodato} G.,  {Testi} L.,
  {P{\'e}rez} L.~M.,  {Bertin} G.,   {Hall} C.,  2021, \mn@doi [\apjl]
  {10.3847/2041-8213/abfe6a}, \href
  {https://ui.adsabs.harvard.edu/abs/2021ApJ...914L..27V} {914, L27}

\bibitem[\protect\citeauthoryear{{Walmswell}, {Clarke}  \&
  {Cossins}}{{Walmswell} et~al.}{2013}]{kicks}
{Walmswell} J.,  {Clarke} C.,   {Cossins} P.,  2013, \mn@doi [\mnras]
  {10.1093/mnras/stt314}, \href
  {https://ui.adsabs.harvard.edu/abs/2013MNRAS.431.1903W} {431, 1903}

\bibitem[\protect\citeauthoryear{{Weidenschilling}}{{Weidenschilling}}{1977}]{metresized}
{Weidenschilling} S.~J.,  1977, \mn@doi [\mnras] {10.1093/mnras/180.2.57},
  \href {https://ui.adsabs.harvard.edu/abs/1977MNRAS.180...57W} {180, 57}

\bibitem[\protect\citeauthoryear{{Youdin}}{{Youdin}}{2011}]{SG11}
{Youdin} A.~N.,  2011, \mn@doi [\apj] {10.1088/0004-637X/731/2/99}, \href
  {https://ui.adsabs.harvard.edu/abs/2011ApJ...731...99Y} {731, 99}

\bibitem[\protect\citeauthoryear{{Youdin} \& {Goodman}}{{Youdin} \&
  {Goodman}}{2005}]{stream1}
{Youdin} A.~N.,  {Goodman} J.,  2005, \mn@doi [\apj] {10.1086/426895}, \href
  {https://ui.adsabs.harvard.edu/abs/2005ApJ...620..459Y} {620, 459}

\bibitem[\protect\citeauthoryear{{Youdin} \& {Johansen}}{{Youdin} \&
  {Johansen}}{2007}]{stream2}
{Youdin} A.,  {Johansen} A.,  2007, \mn@doi [\apj] {10.1086/516729}, \href
  {https://ui.adsabs.harvard.edu/abs/2007ApJ...662..613Y} {662, 613}

\bibitem[\protect\citeauthoryear{{Youdin} \& {Lithwick}}{{Youdin} \&
  {Lithwick}}{2007}]{youdin}
{Youdin} A.~N.,  {Lithwick} Y.,  2007, \mn@doi [\icarus]
  {10.1016/j.icarus.2007.07.012}, \href
  {https://ui.adsabs.harvard.edu/abs/2007Icar..192..588Y} {192, 588}

\makeatother
\end{thebibliography}
\bibliographystyle{mnras}



\onecolumn
\appendix
\section{Dispersion relation}\label{appa}
In this appendix we present the calculations to obtain the dispersion relations for axisymmetric perturbations, with and without taking into account the backreaction.

We consider an infinitesimally thin disc composed of two fluids (gas and dust, henceforth with the subscripts $g$ and $d$). The two components interact through gravitational and drag force. We call $\Sigma_i$ the surface density, $v_i$ the radial velocity, $u_i$ the azimuthal velocity, $h_i$ the enthalpy. In a two-dimensional polar system of coordinates ($r,\phi$), the fluid equations are
\begin{equation}
\partial_{t} \Sigma_{g}+r^{-1} \partial_{r}\left(\Sigma_{g} r v_{g}\right) + \frac{1}{r}\partial_\phi (\Sigma_g u_g)=0 ,
\end{equation}

\begin{equation}
\partial_{t} v_{g}+v_{g} \partial_{r} v_{g}-\frac{u_{g}^{2}}{r} +\frac{u_g}{r}\partial_\phi v_g =-\partial_{r}\left(\Phi+h_{g}\right)+\frac{\Sigma_d}{\Sigma_{g}t_s}\left(v_{d}-v_{g}\right),
\end{equation}

\begin{equation}
\partial_{t} u_{g}+v_{g} \partial_{r} u_{g}- \frac{u_{g} v_{g}} {r} + \frac{u_g}{r}\partial_\phi u_g=-\frac{1}{r}\partial_{\phi}\left(\Phi+h_{g}\right)+\frac{\Sigma_d}{\Sigma_{g}t_s}\left(u_{d}-u_{g}\right),
\end{equation}

\begin{equation}
d h_{g}=c_{g}^{2} \frac{d \Sigma_{g}}{\Sigma_{g}},
\end{equation}

\begin{equation}
\partial_{t} \Sigma_{d}+r^{-1} \partial_{r}\left(\Sigma_{d} r v_{d}\right) + \frac{1}{r}\partial_\phi (\Sigma_d u_d)=0 ,
\end{equation}

\begin{equation}
\partial_{t} v_{d}+v_{d} \partial_{r} v_{d}-\frac{u_{d}^{2}}{ r} + \frac{u_d}{r}\partial_\phi v_d =-\partial_{r}\left(\Phi+h_{d}\right)-\frac{1}{t_s}\left(v_{d}-v_{g}\right) ,
\end{equation}

\begin{equation}
\partial_{t} u_{d}+v_{g} \partial_{r} u_{d}-\frac{u_{d} v_{d}}{ r} + \frac{u_d}{r}\partial_\phi u_d =-\frac{1}{r}\partial_{\phi}\left(\Phi+h_{d}\right)-\frac{1}{t_s}\left(u_{d}-u_{g}\right) ,
\end{equation}

\begin{equation}
d h_{d}=c_{d}^{2} \frac{d \Sigma_{d}}{\Sigma_{d}},
\end{equation}

\begin{equation}
    \nabla^2 \Phi = 4\pi G \delta(z) (\Sigma_g + \Sigma_d).
\end{equation}

The basic state we consider is characterized by uniform surface densities $\Sigma_{g0},\Sigma_{d0}$, azimuthal velocity $u_{g0}=u_{d0} = r\Omega$, zero radial velocity $v_{g0} = v_{d0}=0$ and constant dispersion velocities $c_g,c_d$.

We now perform a first order perturbation analysis of the previous equations: we consider a perturbation to the basic equilibrium state $X_0 + X_1(r,\phi,t)$ that have spatial and temporal dependence like $X_1\propto \exp[i(kr-\omega t + m\phi)]$. 
We focus our attention on axisymmetric perturbations $(m=0)$: we substitute the perturbed quantities into the fluid equations, and we discard any term that is quadratic in them: the six first order perturbed equations are

\begin{equation}
    -i (\omega-m\Omega) \Sigma_{g 1}+i k \Sigma_{g 0} v_{g 1}=0,
\end{equation}

\begin{equation}
    -i (\omega-m\Omega) v_{g 1}-2 \Omega u_{g 1} - \frac{\Sigma_{d0}}{\Sigma_{g0}}\frac{1}{t_s} (v_{d1}-v_{g1})=-\partial_{r}\left(\Phi_{1}+h_{g 1}\right),
\end{equation}

\begin{equation}
    -i (\omega-m\Omega) u_{g 1}-2 B v_{g 1}-\frac{\Sigma_{d0}}{\Sigma_{g0}}\frac{1}{t_s}(u_{d1}-u_{g1})=0,
\end{equation}

\begin{equation}
    -i (\omega-m\Omega) \Sigma_{d 1}+i k \Sigma_{d 0} v_{d 1}=0,
\end{equation}

\begin{equation}
    -i (\omega-m\Omega) v_{d 1}-2 \Omega u_{d 1}+\frac{1}{t_s}\left(v_{d 1}-v_{g 1}\right)=-\partial_{r}\left(\Phi_{1}+h_{d 1}\right),
\end{equation}

\begin{equation}
    -i (\omega-m\Omega) u_{d 1}-2 B v_{d 1}+\frac{1}{t_s}\left(u_{d 1}-u_{g 1}\right)=0,
\end{equation}
where $B(r) = -\frac{1}{2} \frac{\text{d}(\Omega r)}{\text{d}r}+ \Omega$ is the Oort parameter and $4B\Omega^2 = -\kappa^2$. Once we know the form of the density perturbation, it is possible to solve the Poisson equation and get the potential perturbation: it is possible to show that
\begin{equation}
    \Phi_{1}=-\frac{2 \pi G}{|k|}\left(\Sigma_{g 1}+\Sigma_{d 1}\right),
\end{equation}
and the enthalpy can be written as
\begin{equation}
    h_{i, 1}=c_{i}^{2} \frac{\Sigma_{i, 1}}{\Sigma_{i, 0}}.
\end{equation}
Hence, it is possible to write the r.h.s of Euler equations as a function of the perturbed densities
\begin{equation}
    -\partial_r(\Phi_1 + h_{i,1}) = ik\Sigma_{i,1} \left(\frac{2\pi G}{|k|} - \frac{c_i^2}{\Sigma_{i,0}}\right) + ik\frac{2\pi G }{|k|}\Sigma_{j,1}, \quad i\neq j.
\end{equation}

Now, we'll consider the case of axisymmetric perturbations $(m=0)$.

\subsubsection{Without backreaction}
Here, we do not take into account the backreaction, so the drag terms appear only in dust equations. In this case, we can write the matrix of coefficients of $\mathbf{x} = (\Sigma_{g1}, v_{g1},u_{g1},\Sigma_{d1},v_{d1},u_{d1})$ that is
\begin{equation}
A=
\left(\begin{array}{cccccc}
-i\omega & i \Sigma_{g0} k & 0 & 0 & 0 & 0 \\
-i k\left(\frac{2 \pi G}{|k|}-\frac{c_{g}^{2}}{\Sigma_{g 0}}\right) & -i\omega & -2\Omega^2 & -ik\frac{2\pi G}{|k|} & 0 & 0 \\
0 & -2B & -i\omega & 0 & 0 & 0 \\
0 & 0 & 0 & -i\omega & i\Sigma_{d0}k & 0 \\
-ik\frac{2\pi G}{|k|} & -\frac{1}{t_s} & 0 & -i k \left(\frac{2 \pi G}{|k|}-\frac{c_{d}^{2}}{\Sigma_{d 0}}\right) & \frac{1}{t_s}-i\omega & -2\Omega^2 \\
0 & 0 & -\frac{1}{t_s} & 0 & -2B & \frac{1}{t_s}-i\omega
\end{array}\right),
\end{equation}
such that $A\mathbf{x} = 0$. In order to compute the dispersion relation, we impose that the determinant of $A$ is zero: for simplicity, we express the relationship as a function of $y=-i\omega$. To be consistent with \citep{bertinromeo}, we define 
\begin{equation}
    \alpha_i = \kappa^2-\lambda_i = \kappa^2 + c_i^2 k^2 - 2\pi G \Sigma_{i0} |k|,
\end{equation}
and
\begin{equation}
    \beta_i = 2\pi G \Sigma_{i0}|k|,
\end{equation}
and we get
\begin{equation}
\begin{array}{ll}
    y^5 + 2t_s^{-1} y^4 + y^3(\alpha_g+\alpha_d + t_s^{-2}) + t_s^{-1} y^2(2\alpha_g+\alpha_d -\kappa^2-\beta_d) +y [\alpha_g \alpha_d - \beta_g \beta_d + t_s^{-2}(\alpha_g-\beta_d)] + t_s^{-1}[\alpha_g \alpha_d - \beta_g\beta_d - \kappa^2(\alpha_g - \beta_d)]=0 .
\end{array}
\end{equation}

\subsubsection{With backreaction}
Starting from the same basic state as before, now we consider the backreaction, i.e. the effect of the drag force onto the gas component. In this case, the coefficients' matrix is

\begin{equation}
A_\text{B}=
\left(\begin{array}{cccccc}
-i\omega & i \Sigma_{g0} k & 0 & 0 & 0 & 0 \\
-i k\left(\frac{2 \pi G}{|k|}-\frac{c_{g}^{2}}{\Sigma_{g 0}}\right) & \epsilon\frac{1}{t_s} -i\omega & -2\Omega^2 & -ik\frac{2\pi G}{|k|} & -\epsilon\frac{1}{t_s} & 0 \\
0 & -2B & \epsilon\frac{1}{t_s} -i\omega & 0 & 0 & -\epsilon\frac{1}{t_s} \\
0 & 0 & 0 & -i\omega & i\Sigma_{d0}k & 0 \\
-ik\frac{2\pi G}{|k|} & -\frac{1}{t_s} & 0 & -i k \left(\frac{2 \pi G}{|k|}-\frac{c_{d}^{2}}{\Sigma_{d 0}}\right) & \frac{1}{t_s}-i\omega & -2\Omega^2 \\
0 & 0 & -\frac{1}{t_s} & 0 & -2B & \frac{1}{t_s}-i\omega
\end{array}\right),
\end{equation}
where $\epsilon = \Sigma_{d0}/\Sigma_{g0}$. As before, by imposing that the determinant of $A_B$ is zero, we obtain
\begin{equation}
\begin{array}{ll}
    y^{5}+{2t_s^{-1} y^{4}}(1+\epsilon)+ y^{3}\left[ {\alpha}_{g}+ {\alpha}_{d}+t_s^{-2}(1+2\epsilon + \epsilon^2)\right]+{y^{2}}t_s^{-1}\left[(2  {\alpha}_{g}+ {\alpha}_{d}-1)(1+\epsilon)- {\beta}_{d}-\epsilon\beta_g\right]+ \\ 
    +y\left[ {\alpha}_{g}  {\alpha}_{d}- {\beta}_{g}  {\beta}_{d}+ t_s^{-2}\left( {\alpha}_{g}- {\beta}_{d} + \epsilon( {\alpha}_{g}+ {\alpha}_{d}-  {\beta}_g-  {\beta}_d) + \epsilon^2( {\alpha}_{d}- {\beta}_g)    \right)\right]+
    \\ + t_s^{-1}\left[\left( {\alpha}_{g}  {\alpha}_{d}- {\beta}_{g}  {\beta}_{d}\right)(1+\epsilon)-\left( {\alpha}_{g}- {\beta}_{d}\right) - \epsilon\left( {\alpha}_d -  {\beta}_g \right)\right]=0.
\end{array}
\end{equation}




\bsp	
\label{lastpage}
\end{document}